\documentclass[12pt,twoside]{article}

\usepackage{amsmath,amsfonts,amsthm,amssymb,amsopn,amstext,amscd}
\usepackage{enumerate}
\usepackage{graphicx}
\usepackage{float}
\usepackage{xcolor}
\usepackage{longtable}
\usepackage{setspace} 
\usepackage{enumitem}
\usepackage[round]{natbib}
\usepackage{hyperref}
\usepackage{dsfont}
\usepackage{url}
\usepackage{soul}
\usepackage[normalem]{ulem} 
\usepackage{multirow}
\allowdisplaybreaks

\newtheorem{teorema}{Theorem}[section]
\newtheorem{conjetura}{Conjecture}[section]

\newtheorem{proposicion}{Proposition}[section]
\newtheorem{corolario}{Corollary}[section]
\newtheorem{lema}{Lemma}
\newtheorem{nota}{Remark}[section]

\newcommand{\N}{\mathbb{N}}

\newcommand{\tZ}{\widetilde{Z}}

\newcommand{\G}{\mathcal{G}}

\newcommand{\trho}{\tilde{\rho}}
\newcommand{\tm}{\widetilde{m}}
\newcommand{\ts}{\tilde{s}}
\newcommand{\tsigma}{\tilde{\sigma}}

\newcommand{\tT}{\widetilde{T}}
\newcommand{\tF}{\widetilde{F}}
\newcommand{\tM}{\widetilde{M}}
\newcommand{\tp}{\tilde{p}}
\newcommand{\talpha}{\tilde{\alpha}}
\newcommand{\tx}{\tilde{x}}

\newcommand{\tphi}{\tilde{\phi}}

\newcommand{\ind}[1]{\mathds{1}_{\{#1\}}}
\newcommand{\indi}[1]{\mathds{1}_{#1}}

\newcommand*\samethanks[1][\value{footnote}]{\footnotemark[#1]}

\newcommand{\cuad}{\begin{flushright}\vspace{-2ex}$\Box$\vspace{-2ex}\end{flushright}}

\newenvironment{Prf}[1][\unskip]{%
\par
\noindent
{\textbf{Proof of #1}}\newline
\vspace{-2ex}\noindent{}\newline}\cuad

\begin{document}
\pagenumbering{arabic}
\singlespace

\title{A two-sex branching process with oscillations: application to predator-prey systems}
\author{Cristina Guti\'errez\thanks{Both authors contributed equally to this work.}\ \footnote{Department of Mathematics, University of Extremadura, 10071, C\'aceres, Spain. E-mail address: \url{cgutierrez@unex.es}. ORCID: 0000-0003-1348-748X.}  \and Carmen Minuesa\samethanks[1]\  \footnote{Department of Mathematics, Autonomous University of Madrid, 28049, Madrid, Spain. E-mail address: \url{carmen.minuesa@uam.es}. ORCID: 0000-0002-8858-3145.} }
\maketitle

\begin{abstract}
A two-type two-sex branching process is introduced with the aim of describing the interaction of predator and prey populations with sexual reproduction and promiscuous mating. In each generation and in each species the total number of individuals which mate and produce offspring is controlled by a binomial distribution with size given by this number of individuals and probability of success depending on the density of preys per predator. The resulting model enables us to depict 
the typical cyclic behaviour of predator-prey systems under some mild assumptions on the shape of the function that characterises the probability of survival of the previous binomial distribution. We present some basic results about fixation and extinction of both species as well as conditions for the coexistence of both of them. We also analyse the suitability of the process to model real ecosystems comparing our model with a real dataset.
\end{abstract}



\section{Introduction}\label{sec:Introduction}
Recently, the first stochastic process to model the interplay of predator and prey populations with sexual reproduction was introduced in \cite{GutierrezMinuesa2020}. Precisely, the model is a two-type two-sex controlled branching process with a promiscuous mating system and where the control of the populations depends on the current number of individuals of each type in the ecosystem. The model has a recurrent dynamics in each generation where we distinguish three phases: reproduction, control and mating. First, during the reproduction stage couples of each species produce offspring. Next, during the control stage, predators may die due to the lack of preys and preys can be killed by predators. We also allow the possibility of death of individuals by other causes (hunting, existence of other predators, other natural causes, etc.) during this phase. The last stage is the mating phase where females and males of each species mate before giving birth to their children during the reproduction phase at the following generation. Particularly, we consider a promiscuous mating system, that is, where a male may mate with several females, but whenever there is some male in the population each female mates with only one of them. Examples of this kind of mating system in the nature can be found in \cite{Norris1988}, \cite{Furtbauer2011}, \cite{Balme2012}, \cite{Lifjeld2019}, \cite{Wightman2019} or \cite{Lee2019}. In this setting, we establish conditions for the extinction or coexistence of both species and the fixation of one of them. 
The reader is also referred to \cite{GutierrezMinuesa2020} 
for further motivation and background information on the literature on predator-prey models.

Although this paper constitutes the first approach to the problem 
of modelling predator-prey systems with sexual reproduction, the model presents some drawbacks. Namely, the probability of survival of each individual in both populations depends on the sizes of both species in absolute terms. Given that the prey-to-predator ratio is known to be a useful predictor of predation rate and prey growth rate (see \cite{Vucetich-2011}), a more realistic situation is to consider that the survival of each individual depends on this density or proportion of preys per predator and its relation with some parameter $\mu$. Intuitively, this parameter represents the proportion of preys per predator that enables the populations to stay stable and it plays an important role in the description of the behaviour of the process regarding the coexistence of both species, as we state below. 

While this modification retains the generation-by-generation dynamics described above, it leads to a new model that is the focus of this work. An additional advantage of this new process versus the model in \cite{GutierrezMinuesa2020} is that it enables us to model the cyclic behaviour observed in many predator-prey systems. 
We highlight that this process extends the dynamics of other well-known predator-prey models defined via ordinary differential equations (ODEs) and to that end, we examine the similarities between these models and our predator-prey process. We also illustrate these facts through several simulated examples and show the suitability of the model to mimic the evolution of real predator-prey systems. 
Motivated by the impact of predator-prey systems on human life (see, for instance, \cite{Karachle-2016}, \cite{Bouletreau-2018}, or \cite{Ohlberger-2019}) and the possible applicability of our results for the preservation of species, we study the behaviour of our model in relation to the extinction problem. More precisely, we analyse the possible fixation of each species and we provide sufficient conditions for the coexistence of both the predator and prey populations. 




Apart from this introduction, the paper is organised in 6 sections and one appendix. In Section~\ref{sec:Definition} we provide the formal definition of the model and an intuitive interpretation of the assumptions. In Section~\ref{sec:extinction} we study the fixation of each species and the possibility of the ultimate extinction of the whole predator-prey system. Section~\ref{sec:coexistence} is devoted to the analysis of the coexistence of both species. In Section~\ref{sec:real-data} we illustrate the versatility of our process to model real world predator-prey systems. We summarise the main results of this work in Section~\ref{sec:Discussion}. The proofs of all the results are collected in a final appendix to ease the readability of the paper.

In the following, all the random variables (r.v.s) are defined on the same probability space $(\Omega,\mathcal{A},P)$. Moreover, we write $\N_0=\N\cup\{0\}$, and let $\indi{A}$ denote the indicator function of the set $A$.

\section{The probability model}\label{sec:Definition}

In this section, we are concerned with a two-type two-sex and density-dependent branching process and its application to predator-prey systems. This model evolves as a three-stage model of reproduction, control and mating in each generation. We also consider a promiscuous mating as the mating system. We shall start with the formal definition and interpretation of the model.

Let us denote $Z_n$ and $\tZ_n$ the number of predator couples and prey couples at the $n$-th generation, respectively, and assume that the population starts with an arbitrary number of predator and prey couples:
$$(Z_0,\tZ_0)=(z,\tilde{z}) \in \mathbb{N}^2.$$

Let us fix a generation $n\in\N_0$. The dynamics of the three phases in the next generations and the interpretation of the variables involved in the model is as follows. First, in the \textit{reproduction phase}, every couple of each species produces offspring independent of the others and in accordance with the same probability law. Let us denote $\xi_{ni}$ the number of offspring of the $i$-th predator couple at generation $n$ while $\tilde{\xi}_{ni}$ denotes the number of offspring of the $i$-th prey couple at generation $n$, and let us assume that the common probability distributions may vary between the different species, but it remains constant over the generations for each species. Therefore, $\{\xi_{ni}: n\in\N_0, i\in\N\}$ and $\{\tilde{\xi}_{ni}: n\in\N_0, i\in\N\}$ are two independent families  of r.v.s and the r.v.s. within each family are independent and identically distributed non-negative and integer valued. 
At the end of this phase, the sum of all the offspring of each species gives us the total number of predators, $T_{n+1}$, and the total number of preys, $\tT_{n+1}$, in the following generation $n+1$:
\begin{equation}\label{def:model-total-indiv}
\qquad (T_{n+1},\tT_{n+1})=\left(\sum_{i=1}^{Z_{n}}\xi_{ni},\sum_{i=1}^{\tZ_{n}}\tilde{\xi}_{ni}\right),
\end{equation}
where the empty sums are defined as 0.

The reproduction stage is followed by the \textit{control phase}, where the number of predators and preys that are available for reproduction might not coincide with the total number of individuals due to several reasons (hunting, lack of food supply, their capture by predators, reproduction disability, etc.) and therefore they might be reduced. Thus, if there are $T_{n+1}=t$ predators and $\tT_{n+1}=\tilde{t}$ preys in the population, the r.v.s $\phi_{n+1}(t,\tilde{t})$ and $\tilde{\phi}_{n+1}(t,\tilde{t})$ denote the number of predators and preys that survive and are able to reproduce in generation $n+1$, respectively. Then, we assume that $\{\phi_{n}(t,\tilde{t}): n,t,\tilde{t}\in\N_0\}$ and $\{\tphi_{n}(t,\tilde{t}): n,t,\tilde{t}\in\N_0\}$ are two independent sequences of independent and non-negative integer valued r.v.s. We also consider that the survival of each predator (prey) is independent of the generation and of the survival of the remaining predators (preys), and that the probability of survival is the same for all the individuals in the same species. Consequently, it is natural to assume binomial distributions for the control variables. Precisely, if there are $t\in\N$ predators and $\tilde{t}\in\N$ preys in the population at generation $n+1$, then we assume that the distribution of the variable $\phi_{n+1}(t,\tilde{t})$ is binomial with size $t$ and probability of success $s(\tilde{t}/t)$, whereas the one of $\tphi_{n+1}(t,\tilde{t})$ is also binomial with size $\tilde{t}$ and probability $\tilde{s}(\tilde{t}/t)$, with $s,\ts:[0,\infty)\to [0,1]$ continuous functions, that is, for $t,\tilde{t}\in\N$,
\begin{align*}
\phi_{n+1}(t,\tilde{t})\sim B(t,s(\tilde{t}/t)),\quad \tphi_{n+1}(t,\tilde{t})&\sim B(\tilde{t},\ts(\tilde{t}/t)).
\end{align*}
We note that $s(\tilde{t}/t)$ and $\tilde{s}(\tilde{t}/t)$ represent the probabilities that a predator and a prey survive, respectively, given that there are $t$ predators and $\tilde{t}$ preys in the population, and these probabilities depend on the density of preys per predator, $\tilde{t}/t$.

At the end of this control phase, there are $F_{n+1}$ females and $M_{n+1}$ males within the survivor predator population, and $\tF_{n+1}$ females and $\tM_{n+1}$ males within the survivor prey population at generation $n+1$ and all of them are able to reproduce. We write $\alpha$ and $\talpha$, with $0<\alpha, \talpha<1$, to denote the probability that a survivor predator and prey is female, respectively. Thus, if there are $k$ predators and $\tilde{k}$ preys that survive the control phase, then the random vector $(F_{n+1},M_{n+1})$ follows a multinomial distribution with parameters $k$ and $(\alpha,1-\alpha)$, and $(\tF_{n+1},\tM_{n+1})$ follows a multinomial distribution of parameters $\tilde{k}$ and $(\tilde{\alpha},1-\tilde{\alpha})$. Formally, conditional on $\{\phi_{n+1}(T_{n+1},\tT_{n+1})=k,\tphi_{n+1}(T_{n+1},\tT_{n+1})=\tilde{k}\}$,
\begin{align*}
(F_{n+1},M_{n+1}) \sim M(k;(\alpha,1-\alpha)),\quad (\tF_{n+1},\tM_{n+1}) \sim M(\tilde{k};(\talpha,1-\talpha)).
\end{align*}

The last step is the \textit{mating phase}, where female and male predators and preys mate in order to form couples at generation $n+1$, $Z_{n+1}$ and $\tZ_{n+1}$. The total number of couples of each species is determined by means of a promiscuous mating function ($f(x,y)=x\min\{1,y\}$) depending on the number of females and males of each species at the current generation:
\begin{align*}\label{def:model-couples}
(Z_{n+1},\tZ_{n+1})&=\left(F_{n+1}\min\{1,M_{n+1}\},\tF_{n+1}\min\{1,\tM_{n+1}\}\right).
\end{align*}

The process $\{(Z_n,\tZ_n)\}_{n\in \mathbb{N}_0}$ defined above is named \emph{predator-prey two-sex and density-dependent branching process} (PP-2SDDBP).
Note that by the definition of the model it is not difficult to verify that the number of predator and prey couples in a certain generation only depends on the total number of predator and prey couples in the previous generation. Thus, the bivariate sequence $\{(Z_n,\tZ_n)\}_{n\in\N_0}$ is a discrete time homogeneous Markov chain whose states are two-dimensional vectors with non-negative integer coordinates. Moreover, since the empty sum is assumed to be zero, if there is no couple of one of the species in some generation, from this generation on, couples and individuals of that species no longer exist. That is, if $Z_n=0$ ($\tZ_n=0$, resp.) for some $n\in\N$, then $T_{k}=0$ ($\tilde{T}_k=0$, resp.) and $Z_k=0$ ($\tZ_k=0$, resp.) for all $k\geq n$. Bearing this in mind, it is not difficult to deduce that $(0,0)$ is an absorbing state and every non-null state is transient.

\medskip

Finally, let us introduce the notation for the distribution of the offspring variables. We write $p_k=P(\xi_{01}=k)$, and $\tp_k=P(\tilde{\xi}_{01}=k)$, for $k\in\N_0$. The distributions $p=\{p_k\}_{k\in \mathbb{N}_0}$ and $\tilde{p}=\{\tilde{p}_k\}_{k\in \mathbb{N}_0}$ are called \textit{offspring distribution} or \emph{reproduction law of the predator and prey populations}, respectively. In order to avoid trivialities, we assume that $p_0+p_1+p_2<1$ and $\tilde{p}_0+\tilde{p}_1+\tilde{p}_2<1$. Moreover, we assume that these distributions have finite and positive means and variances, which are denoted $m$ and $\sigma^2$, respectively, for the predators, and $\tilde{m}$ and $\tilde{\sigma}^2$ for the preys.

\bigskip

A natural question that arises in this setting is what conditions the functions $s(\cdot)$ and $\ts(\cdot)$ should satisfy so that the model can describe realistic situations observed in nature and which are also covered in other predator-prey models, such as, for instance:
\begin{enumerate}[label=(S\arabic*),ref=(S\arabic*)]
\item The limited appetite of predators (see \cite{Arditi-1978}).\label{sit: limit appetite}
\item The survival of predators in absence of preys due to alternative food supplies or their artificial feeding (see \cite{Adamu-2018} or \cite{prasad-2013}).\label{sit:predators survival}
\item The fact that preys have a chance to hide away from predators (see \cite{Gal-2015}).\label{sit:prey escape}
\item The obvious fact that animals die for natural reasons, that is, not all predators die because of the lack of preys and not all preys are killed by predators.\label{sit: die of predators and preys}
\item The typical cyclic behaviour in this type of predator-prey systems given by the fact that when the proportion of preys per predator is small (big) the number of preys and predators should drop (arise)  in the next generations.\label{sit:cycle behavior}
\end{enumerate}

To this end, henceforth we assume that these functions satisfy the next conditions for certain constants $0<\rho_1<\rho_2<1$, $0<\trho_1<\trho_2<1$ and $\mu>0$:
\begin{enumerate}[label=(C\arabic*),ref=(C\arabic*)]
\item  $s(\cdot)$ and $\ts(\cdot)$ are strictly increasing continuous functions.\label{cond: s and ts continuity}
\item $\lim_{x\to \infty}s(x)=\rho_2$ and $\lim_{x\to \infty}\ts(x)=\trho_2$.\label{cond: s y ts limits in infty}
\item $\lim_{x\to 0}s(x)=\rho_1$ and $\lim_{x\to 0}\ts(x)=\trho_1$.\label{cond: s y ts limits in 0}
\item The r.v. $\phi_n(k,0)$ follows a binomial distribution with parameters $k$ and $\rho_1$, and $\tphi_n(k,0)=0$ a.s., for each $n\in\N$, and $k\in\N_0$. \label{cond:0-preys}
\item The r.v. $\tphi_n(0,\tilde{k})$ follows a binomial distribution with parameters $\tilde{k}$ and $\trho_2$, and $\phi_n(0,\tilde{k})=0$ a.s., for each $n\in\N$, and $\tilde{k}\in\N_0$. \label{cond:0-predators}
\item $s(\mu)=\displaystyle\frac{1}{\alpha m}$ and $\tilde{s}(\mu)=\displaystyle\frac{1}{\talpha \tm}$.\label{cond: s y st in mu}
\end{enumerate}

The PP-2SDDBP $\{(Z_n,\tZ_n)\}_{n\in \mathbb{N}_0}$ satisfying conditions \ref{cond: s and ts continuity}-\ref{cond: s y st in mu} is called \emph{predator-prey two-sex and density-dependent branching process with oscillations} (PP-2SDDBPO). The intuitive interpretation of those conditions is now explained.

The monotonicity property (condition \ref{cond: s and ts continuity}) of $s(\cdot)$ and $\ts(\cdot)$ is a natural assumption since if the ratio of the preys to the predators is very large, then there are enough resources for the predators to survive and many preys also survive whenever we assume the limited appetite of predators (situation \ref{sit: limit appetite}) together with the possibility  that preys escape from predators (situation \ref{sit:prey escape}). Consequently, the probability of survival of each predator and prey should be high. The greater this ratio becomes, the greater the probability of survival of each predator and prey. This occurs because the available resources  of each predator -in relative terms- increases so does the probability of predator survival. Moreover, each prey has a less chance to be captured by predators, and as a result the probability that a prey survives arises.

Condition \ref{cond: s y ts limits in infty} (\ref{cond: s y ts limits in 0}, resp.) means that when the number of preys is much greater (smaller) than the number of predators in some generation, the probability of survival of predators and preys converge to certain constants $\rho_2$ and $\trho_2$ ($\rho_1$ and $\trho_1$). Intuitively, $\rho_1$ is the survival probability of predators when there is no prey and $\trho_2$ is the survival probability of preys when there is no predator. Moreover:
\begin{itemize}
\item The condition $\rho_1>0$ covers situation \ref{sit:predators survival} and it means that a predator can survive although the proportion of preys per predator goes to 0 due to the availability of other food resources.

\item The fact that preys could escape from predators, situation \ref{sit:prey escape}, is described by $\trho_1>0$ because this means that a prey can survive even if the proportion of preys per predators converges to 0.

\item The assumption $\rho_2<1$ represents situation \ref{sit: die of predators and preys}, that is, not all predators might be available to form couples even if there are enough preys in the population because some of them could die by different reasons as natural death, hunting or their own predators.

\item Similarly, the constraint $\trho_2<1$ indicates that even in absence of predators  the preys could die by other causes
, and thus it describes the situation \ref{sit: die of predators and preys}.
\end{itemize}

To cover situations \ref{sit: limit appetite} and \ref{sit:cycle behavior}, we introduce the parameter $\mu$ interpreted as the necessary prey density per predator so that both populations stay stable. 
Suppose that there are $t$ predators and $\tilde{t}$ preys in the population in certain generation. When the proportion of preys per predator is less than  
prey density needed to maintain the predator population, that is, $\tilde{t}/t<\mu$, then there is a lack of preys to feed the predators and some of them may die of starvation, thus, the expected number of predators at the next generation should be less than the current population size. The same behaviour should occur for the prey population because the chance to be captured by the predators is greater as a result of the large number of predators. Then, in this case, the number of individuals of both populations should decrease. On the other hand, when the proportion of preys per predator is greater than 
prey density needed to maintain the predator population around the same population size, that is, $\tilde{t}/t>\mu$, then the predators have enough preys to feed themselves and survive and consequently, the expected number of predators or preys at the next generation should be greater than the current number of those individuals in the ecosystem.

The above facts are described through condition \ref{cond: s y st in mu}. For a better understanding of the meaning this condition we establish the following proposition. The proof is analogous to the one of Proposition 7~\emph{(i)} in \cite{GutierrezMinuesa2020} and it is therefore omitted. Let us first introduce the following notation:
\begin{align*}
\mathcal{G}_n&=\sigma(T_{l},\tilde{T}_{l}:l=1,\ldots,n),\quad n\in\N.
\end{align*}
\begin{proposicion}\label{prop:expectations}
Let $\{(Z_n,\tZ_n)\}_{n \in \mathbb{N}_0}$ be a PP-2SDDBPO and let us fix $n\in\N$. Then: 
\begin{enumerate}[label=(\roman*),ref=\emph{(\roman*)}]
\item The conditional expectations of the number of individuals of each species are\label{prop:expectations-i}
\begin{align*}
E[T_{n+1}|\mathcal{G}_n]&=\alpha m T_ns(\tT_n/T_n)\Big[1-\big(1-s(\tT_n/T_n)(1-\alpha)\big)^{T_n-1}\Big]\ind{T_n>0},\\
E[\tT_{n+1}|\mathcal{G}_n]&=\talpha \tm \tT_n\ts(\tT_n/T_n)\Big[1-\big(1-\ts(\tT_n/T_n)(1-\talpha)\big)^{\tT_n-1}\Big]\ind{T_n>0}\\
&\phantom{=}+\talpha \tm \tT_n\trho_2\Big[1-\big(1-\trho_2(1-\talpha)\big)^{\tT_n-1}\Big]\ind{T_n=0}.
\end{align*}

\item The conditional variances of the number of individuals of each species satisfy:\label{prop:expectations-ii}
\begin{align*}
Var[T_{n+1}|\mathcal{G}_n]&\leq \Big[2\alpha^2 m^2 T_n^2 s(\tT_n/T_n)^2 (1-s(\tT_n/T_n)(1-\alpha))^{T_n-1}\\
&\phantom{\leq} +(\sigma^2+m^2)\alpha T_ns(\tT_n/T_n)\Big]\ind{T_n>0},\\
Var[\tT_{n+1}|\mathcal{G}_n]&\leq 2\talpha^2 \tm^2\tT^2_n \Big[\ts(\tT_n/T_n)^2(1-\ts(\tT_n/T_n)+\talpha \ts(\tT_n/T_n))^{\tT_n-1}\ind{T_n>0}\\
&\phantom{=}+\trho_2^2(1-\trho_2+\talpha \trho_2)^{\tT_n-1}\ind{T_n=0}\Big]\\
&\phantom{=}+(\tsigma^2+\tm^2)\talpha \tT_n \Big[\ts(\tT_n/T_n)\ind{T_n>0}+\trho_2\ind{T_n=0}\Big].
\end{align*}
\end{enumerate}
\end{proposicion}

For an intuitive explanation of condition \ref{cond: s y st in mu} let us analyse Proposition \ref{prop:expectations}~\ref{prop:expectations-i}. Let us focus on the predator population and consider that both populations have enough number of individuals at generation $n$ to avoid extinction at the next generation. In particular, $T_n>1$, and $\tT_n>1$. Since $\rho_1>0$, then $1-s(\tT_n/T_n)(1-\alpha)\leq 1-\rho_1(1-\alpha)<1$. Consequently, if $T_n$ becomes very large, then the term $T_n\big(1-s(\tT_n/T_n)(1-\alpha)\big)^{T_n-1}$ is negligible and essentially,
$$E[T_{n+1}|\mathcal{G}_n]\approx\alpha m T_ns(\tT_n/T_n).$$
Now, if the proportion of preys per predator $\tT_n/T_n$ is equal to $\mu$, then it is natural to expect that the number of predators in the generation $n+1$ is roughly the same that in the current generation $n$, and this can be achieved by setting $s(\mu)=1/(\alpha m)$. Under this condition, if $\tT_n/T_n>\mu$, then the expected number of individuals is greater than the current number of predators in the population, whereas if $\tT_n/T_n<\mu$, then there is a drop in the expected number of individuals. The same arguments provide the interpretation for the prey population.



Notice that \ref{cond: s and ts continuity}-\ref{cond: s y ts limits in 0} together with \ref{cond: s y st in mu} imply: 
\begin{equation}\label{eq: rho1alfam<1<rho2alfam}
\rho_1 \alpha m<1<\rho_2 \alpha m \quad \mbox{ and } \quad \trho_1 \talpha \tm<1<\trho_2 \talpha \tm,
\end{equation} 
and therefore $\alpha m>1$ and $\talpha \tm>1$. For this reason, throughout the paper we assume that the parameters of the model satisfy \eqref{eq: rho1alfam<1<rho2alfam}.

\begin{nota}\label{nota:descrip-def}
We provide two examples of functions $s(\cdot)$ and $\tilde{s}(\cdot)$ satisfying \ref{cond: s and ts continuity}-\ref{cond: s y ts limits in 0}. For instance, for the function $s(\cdot)$, we could take
\begin{align*}
f_1(x)=(\rho_2-\rho_1)(1-a^{-x})+\rho_1, \mbox{ with } a>1, \quad \mbox{ or }\quad f_2(x)=\frac{\rho_2x^b+\rho_1}{x^b+1},\mbox{ with } b>0.
\end{align*}
In order to those functions satisfy condition \ref{cond: s y st in mu}, we should take
\begin{align*}
a=\left(\frac{(\rho_2-\rho_1)\alpha m}{\rho_2 \alpha m -1}\right)^{1/\mu},\quad \text{ and }\quad b=\frac{\log((1-\rho_1\alpha m)/(\rho_2 \alpha m -1))}{\log (\mu)},
\end{align*}
whenever $b>0$. Note that $a>0$ always holds under condition \eqref{eq: rho1alfam<1<rho2alfam}.

\end{nota}

\medskip

\subsection{Illustrative examples}\label{subsec:illustrative-examples}


In this subsection, we present two simulated examples to illustrate different behaviours that can be observed in the model. We consider a PP-2SDDBPO with Poisson distributions for both the predator and prey reproduction laws. We simulated the first $n$ generations of this model starting with $Z_0=10$ predator couples and $\tZ_0=10$ prey couples. We take $\mu=2$ in both examples. The remaining values of the parameters are shown in Table~\ref{table: parameter values}.

\begin{table}[H]
\begin{center}
\begin{tabular}[H]{|c|c|c|c|c|c|c|c|c|c|}
\hline 
 & $n$ &$\alpha$ & $m$ & $\rho_1$ & $\rho_2$ & $\talpha$ & $\tm$ & $\trho_1$ & $\trho_2$ \\ 
\hline 
Example 1 & 75 & 0.42 & 3 & 0.45 &0.9  & 0.49 & 4 & 0.25 & 0.6 \\ 
\hline 
Example 2 & 3000 & 0.49 & 4 & 0.25 & 0.6 & 0.42 & 3 & 0.45 & 0.9\\ 
\hline
\end{tabular}
\end{center}
\caption{Values of the parameters of the models in the simulated examples.}\label{table: parameter values}
\end{table} 

\begin{figure}[H]
\centering\includegraphics[width=0.32\textwidth]{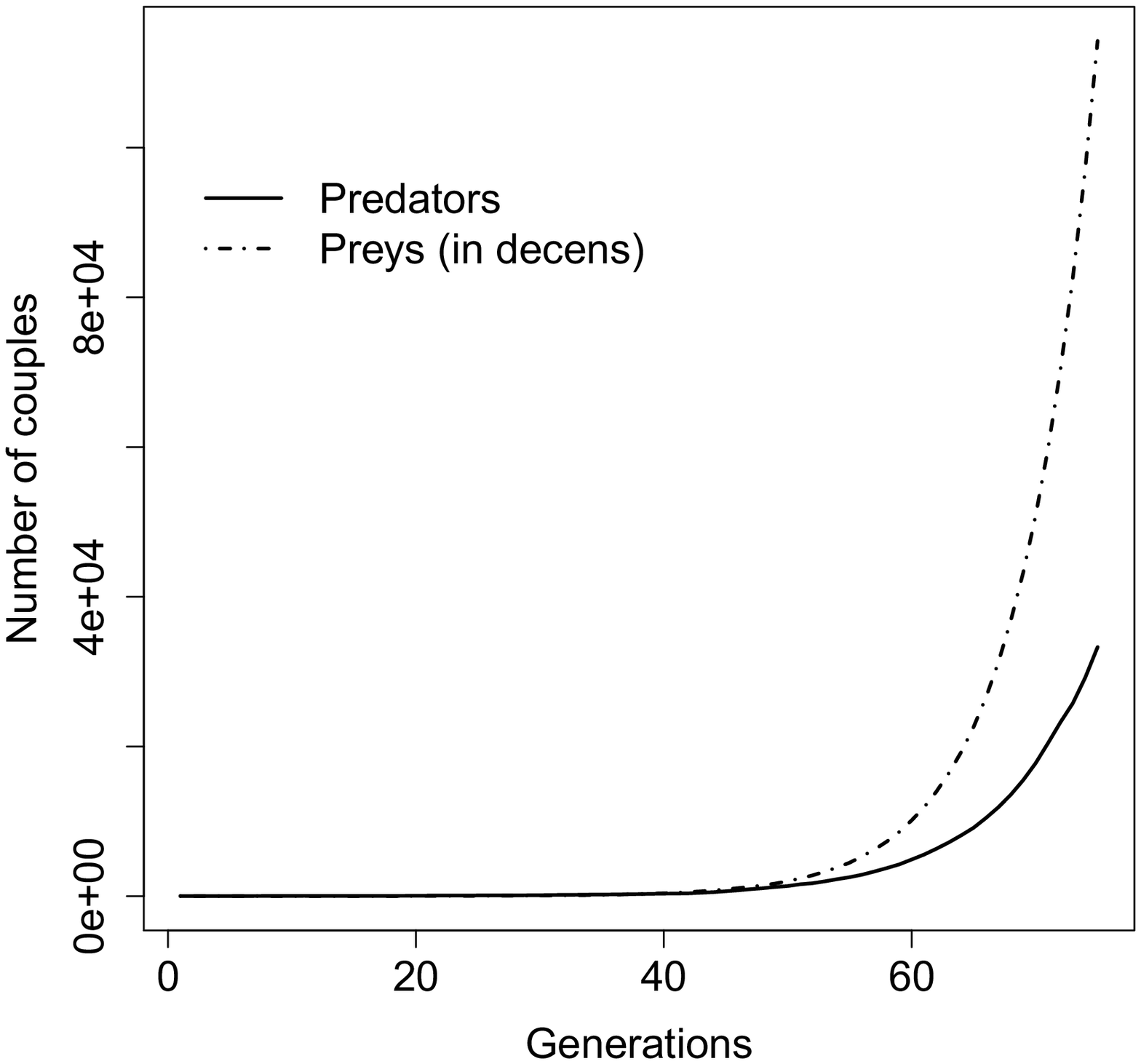}
\includegraphics[width=0.32\textwidth]{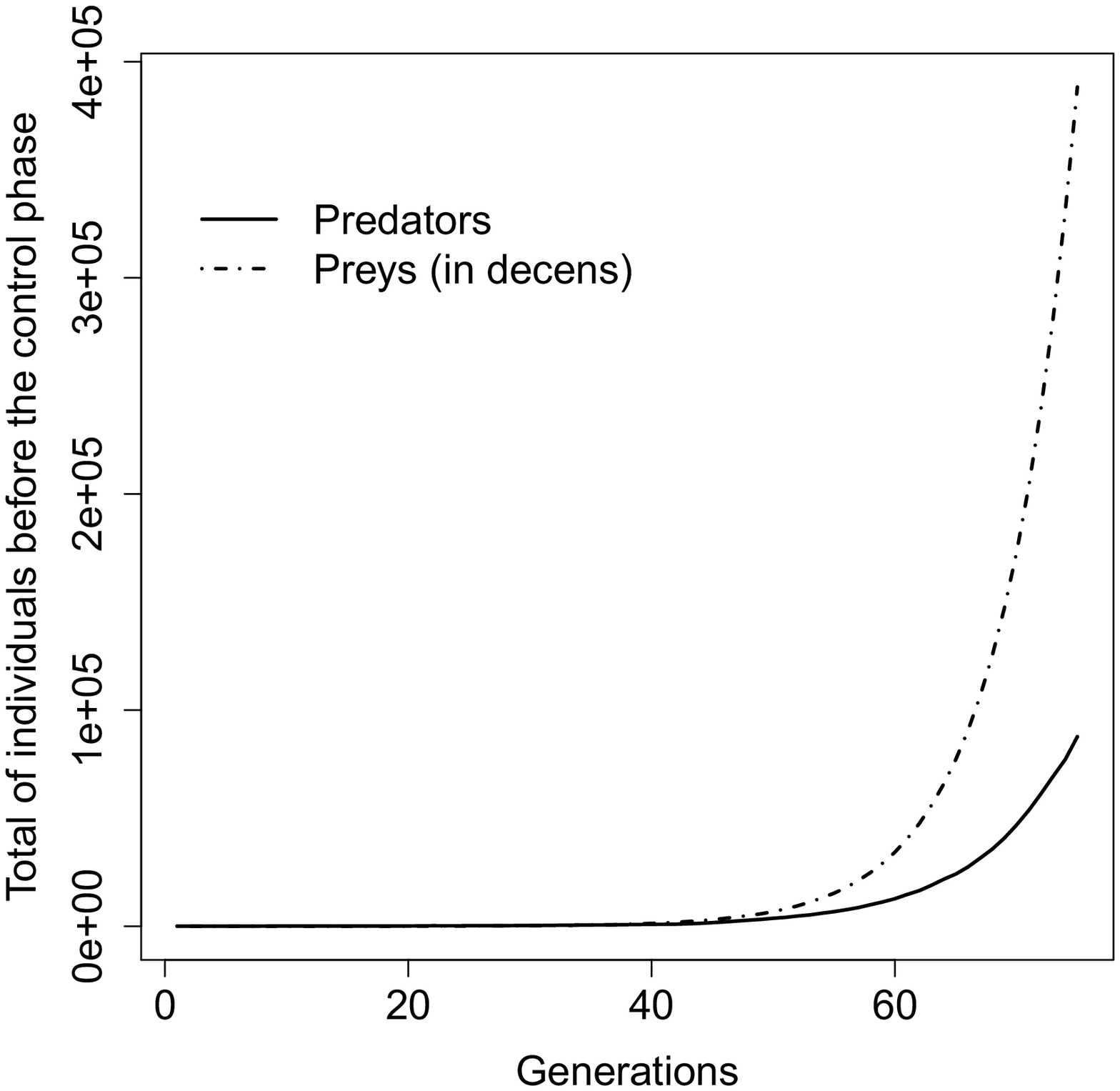}
\includegraphics[width=0.32\textwidth]{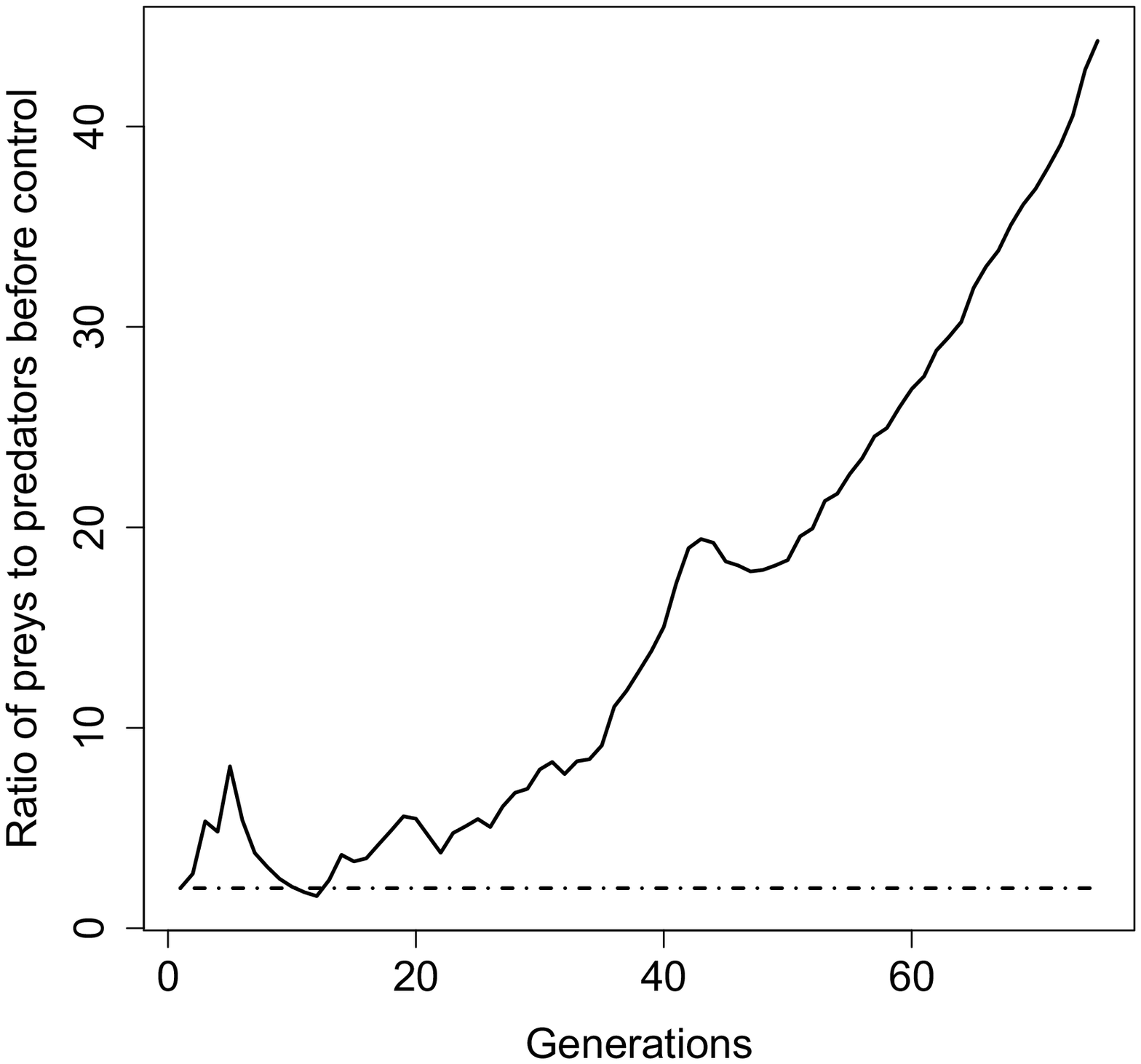}
\caption{\textbf{Example 1}. Left: evolution of the number of predator couples (solid line) and prey couples (dashed-dotted line). Centre: evolution of the total number of predators (solid line) and of the total number of preys (dashed-dotted line) before the control phase. Right: evolution of the ratio of the total of preys to the total of predators before the control phase (black line). Horizontal dashed-dotted line represents the value of $\mu$.}\label{fig:path-density-sit1}
\end{figure}

In Example 1, we have $\trho_2\talpha\tm=1.176>\rho_2\alpha m=1.134$. This can be considered a case where the prey is the dominant in the sense that, while both population are alive, the total number of preys is greater than the total number of predators from one generation onwards. The evolution of the number of couples, the total number of individuals of each species, and the ratio of the total of preys to the total of predators before the control phase are plotted in Figure \ref{fig:path-density-sit1}. There is a clear exponential growth at different regimes in the first two graphs. The last one shows an initial and very short oscillation of the ratio around $\mu$ before its fast increase. A theoretical justification for these results is given in Theorem \ref{thm:coexistence-positive} in Section \ref{sec:coexistence}.

\begin{figure}[H]
\centering\includegraphics[width=0.32\textwidth]{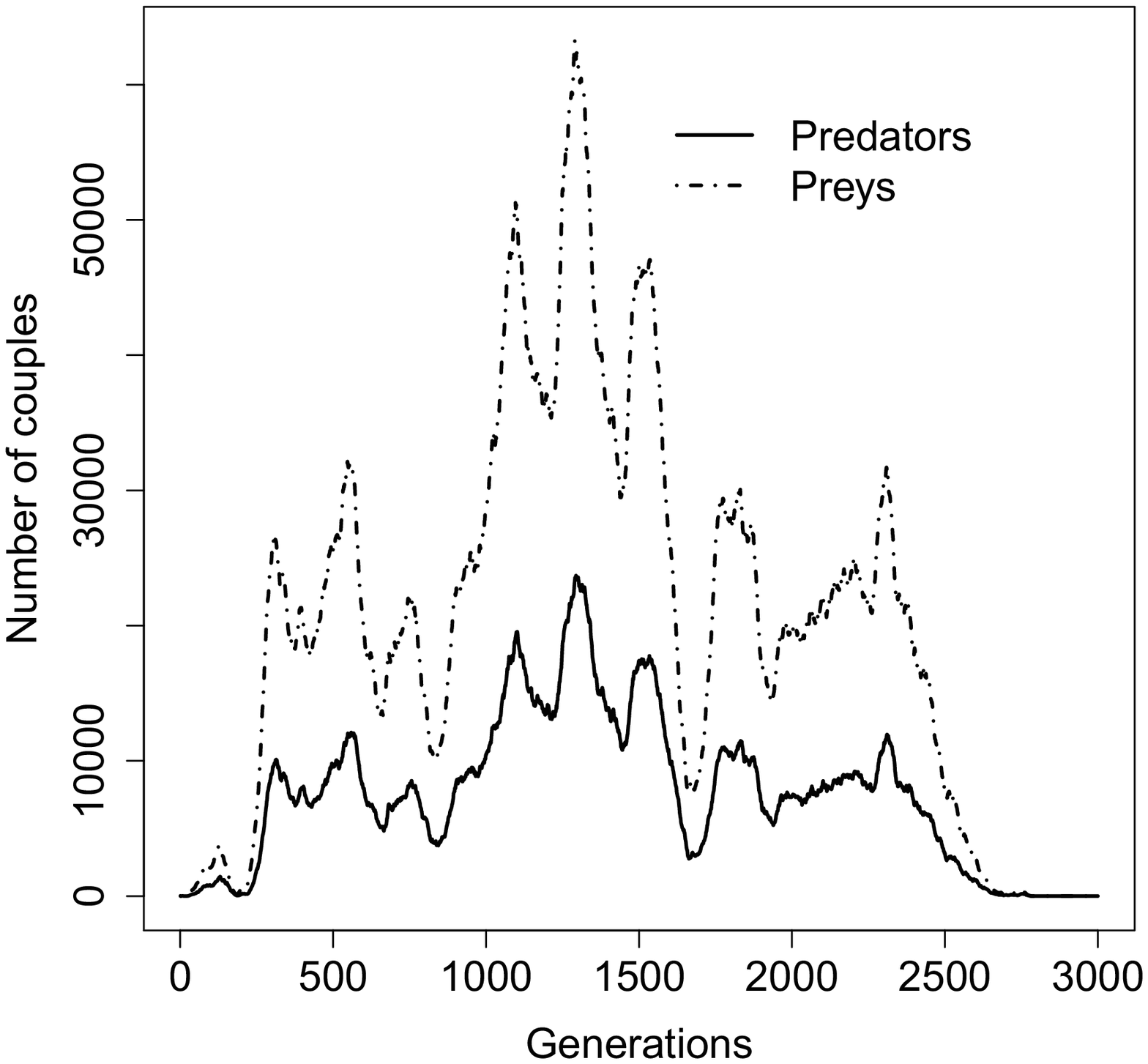}
\includegraphics[width=0.32\textwidth]{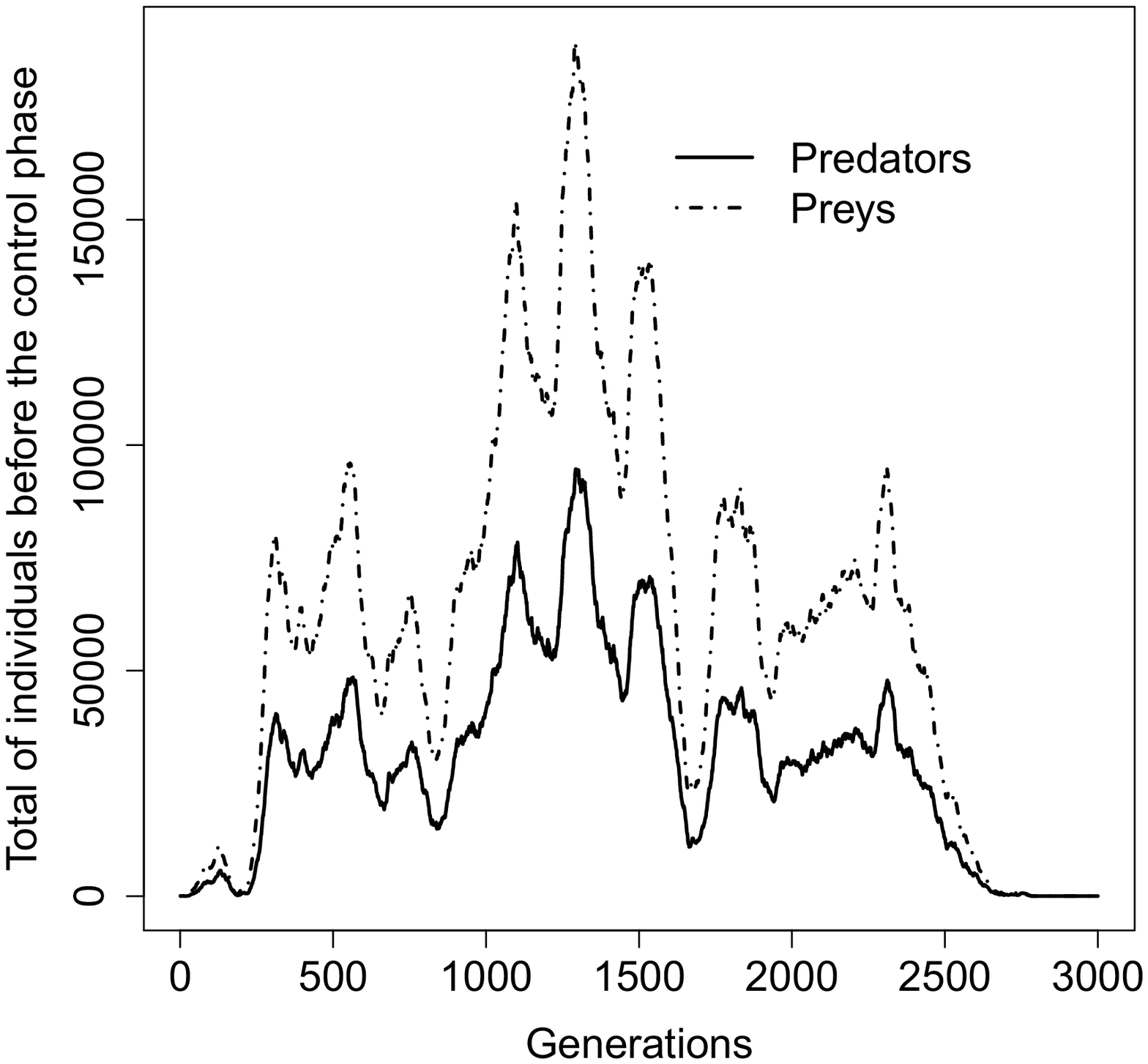}
\includegraphics[width=0.32\textwidth]{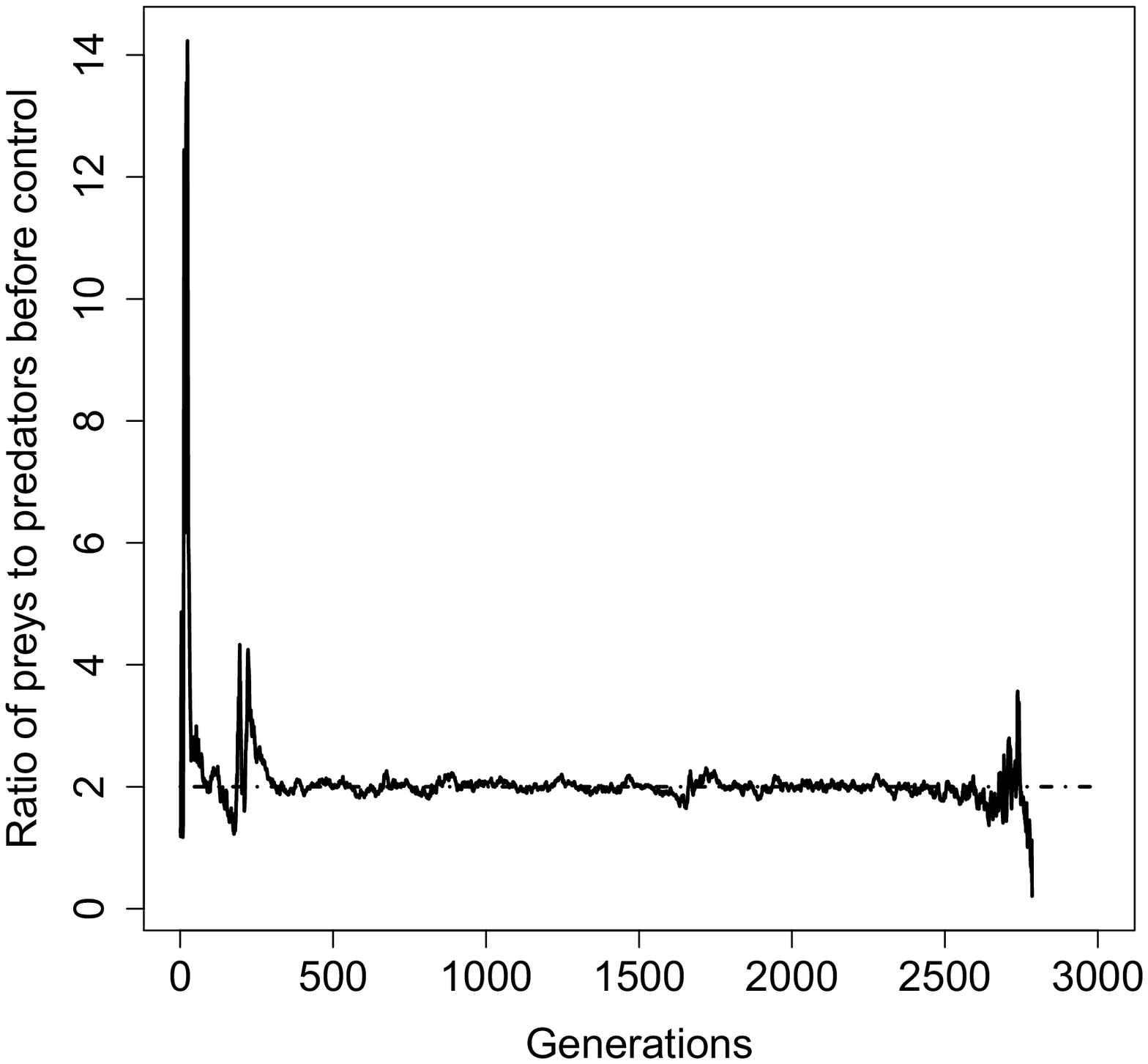}
\caption{\textbf{Example 2.} Left: evolution of the number of predator couples (solid line) and prey couples (dashed-dotted line). Centre: evolution of the total number of predators (solid line) and of the total number of preys (dashed-dotted line) before the control phase. Right: evolution of the ratio of the total of preys to the total of predators before the control phase (black line). Horizontal dashed-dotted line represents the value of $\mu$.}\label{fig:path-density-sit2}
\end{figure}

In Example 2, we show the case when $\rho_2\alpha m=1.176>\trho_2\talpha \tm=1.134$. Although initially one could think that in this case the predators should dominate, what one observes is that the number of preys is greater than the number of predators in almost all generations where both coexist. However, the asymptotic behaviour of the process is completely different from the one in Example 1. To show this fact, we represented the number of couples, the total number of individuals of each species, and the ratio of the total of preys to the total of predator before the control phase over the generations in Figure \ref{fig:path-density-sit2}. As in the previous example, we see an initial exponential growth at different rates in the first two plots until some generation where we note a clear change in the limiting behaviour compared to the Example 1. More precisely, there are some  small and recurrent fluctuations in the number of individuals and couples of each species before the prey population gets extinct at generation 2785. As established in our Proposition~\ref{prop: prey and predator fixation} below, this fact dooms the system to its ultimate extinction as our graphs show. Moreover, a closer look to the graphs shows that an increment in the number of preys is usually followed by an increment on the number of predators while the increment in the number of predators is followed by a drop in the number of preys. Turning to the graph of the ratio of preys to predators, one can observe a totally dissimilar behaviour compared to the one in Example 1. Indeed, we see that the ratios eventually stabilise and oscillate around $\mu$ before they hit the value 0 at generation 2785.

\subsection{Relationship with other predator-prey models}\label{subsec:relation with other models}

Despite the existence of the mating phase, we now make a comparison between our model and the equivalent deterministic model defined by the Lotka-Volterra (predator-prey) equations. To that end, we analyse the evolution of the expected number of individuals of each species that survive in presence of the other one. We shall start with the prey population. 

Let us assume that there are $x$ predators and $\tx$ preys in the ecosystem, then the mean number of preys that are alive after the control phase is $\tilde{x}\ts(\tilde{x}/x)=\tilde{x}-\tilde{x}\tilde{d}(\tilde{x}/x)$, with $\ts(\cdot)=1-\tilde{d}(\cdot)$, and $\tilde{d}(\cdot)$ denoting the probability that a prey does not survive. Thus, the term $\tilde{x}$ corresponds to the growth of the number of preys in the Lotka-Volterra equations whereas $\tilde{x}\tilde{d}(\tilde{x}/x)$ -expected number of preys died during the control- is the counterpart of the interaction term in the Lotka-Volterra equations. We also note that growth rate for the preys in the Lotka-Volterra equations is introduced in our model via the reproduction phase of this species. Moreover, if at some generation there is no predator, then by Proposition~\ref{prop:expectations}
$$E[\tT_{n+1}|T_n=0,\tT_n= \tx]=\talpha \tm \tx\trho_2\Big[1-\big(1-\trho_2(1-\talpha)\big)^{\tx-1}\Big],\quad \tx\in\N,$$
which implies an exponential growth of the prey population  as occurs in predator-prey system of ODEs. 

Next, we examine the mean number of predators which are alive after the control phase.  
This quantity is $xs(\tilde{x}/x)$, which corresponds to the interaction term in the Lotka-Volterra equations. The term corresponding to the death of predators by other causes different from the lack of preys is introduced in our model by the consideration that $\rho_2<1$ in the control phase (see situation \ref{sit: die of predators and preys}).

Now, we evaluate the presence of two key elements of every predator-prey ODEs system: the \emph{predation rate} (PR) and the \emph{kill rate} (KR). The former represents the proportion of the prey population killed by the predators and the latter is the number of preys killed per predator during a unit of time. Since the present model is a discrete-time stochastic process, these quantities refer to expected quantities in each generation. Thus, the equivalent of the predator rate in our model is the probability of survival $\tilde{d}(\tx/x)$, whereas the counterpart of the kill rate is $\tilde{d}(\tx/x) \tx/x$, that is, the expected number of preys caught per predator during the control phase. We emphasise the fact that these quantities satisfies the common relationship
$$PR=\frac{KR\cdot Pd}{Pr},$$
where $Pd$ stands for the number of predators and $Pr$ for the number of preys.

Finally, we remark that on the one hand, the fact that $(0,0)$ is an absorbing state is the equivalent in our model of the fixed point $(0,0)$ for the Lotka-Volterra system, which  represents the ultimate extinction of both populations. On the other hand, 
the parameter $\mu$ in the PP-2SDDBPO (recall that $\mu$ is the prey-to-predator ratio that ensures that both populations stay stable) is the counterpart of the second steady state in the Lotka-Volterra equations, where both populations support their current size (see \cite{Hirsch-Smale-1974}, pp.~258-263 for further details). Moreover, the introduction of the parameter $\mu$ together with the shape of the functions $s(\cdot)$ and $\ts(\cdot)$ enables us to model the fluctuations usually observed in these dynamical systems, as we showed in Example 2 in the previous subsection.

\section{Some basic results on fixation and extinction}\label{sec:extinction}

In this section we examine the evolution of the species in relation to their fixation and extinction. For this purpose, we first consider the following sets: $\{Z_n\to 0,\tZ_n\to 0\}$, termed extinction of predators and prey populations; $\{Z_n \to\infty,\tZ_n\to \infty\}$, coexistence of both species; $\{Z_n\to \infty,\tZ_n\to 0\}$, predator population fixation; and $\{Z_n \to 0,\tZ_n\to\infty\}$, prey population fixation.

In a first result we establish the usual property of branching processes known as the extinction-explosion. The proof of this result is similar to the proof of Proposition 8 in \cite{GutierrezMinuesa2020} and  it is therefore omitted.

\begin{proposicion}\label{prop:extexp}
Let $\{(Z_n,\tZ_n)\}_{n\in\N_0}$ be a PP-2SDDBPO. Then:
\begin{enumerate}[label=(\roman*),ref=\emph{(\roman*)}]
\item $P(\liminf_{n\to\infty}(Z_n,\tZ_n)=(k,\tilde{k}))=0$, and $P(\limsup_{n\to\infty}(Z_n,\tZ_n)=(k,\tilde{k}))=0$, for each $(k,\tilde{k})\in\N_0^2\backslash\{(0,0)\}$.\label{prop:extexp-i}
\item $P(Z_n\to 0,\tZ_n\to 0)+ P(Z_n \to\infty,\tZ_n\to \infty)+P(Z_n\to \infty,\tZ_n\to 0)+
P(Z_n \to 0,\tZ_n\to\infty)=1$.\label{prop:extexp-ii}
\end{enumerate}
\end{proposicion}



Taking into account this result, we examine the behaviour of both species in relation to their extinction. To that end, given $i,j>0$, henceforth we write $P_{(i,j)}(\cdot)=P(\cdot|(Z_0,\tZ_0)=(i,j))$.

\bigskip


First, we study the probability of fixation of each species. In the fixation events, from some generation on, the survivor species behaves as the two-sex branching process with random control on the total number of individuals introduced in Section 2 in \cite{GutierrezMinuesa2020}. The control variables of the corresponding processes follow binomial distributions with constant probability of success $\rho_1$, in the case of the predator fixation, and with probability $\trho_2$ in the case of the prey fixation. Thus, bearing in mind condition \eqref{eq: rho1alfam<1<rho2alfam} the following result immediately holds applying Theorems~1 and 2~(ii)~(b) in the aforementioned paper.

\begin{proposicion}\label{prop: prey and predator fixation}
Let $\{(Z_n,\tZ_n)\}_{n\in\N_0}$ be a PP-2SDDBPO. For any initial values $i,j>0$:
\begin{enumerate}[label=(\roman*),ref=\emph{(\roman*)}]
\item $P_{(i,j)}(Z_n\to \infty,\tZ_n\to 0)=0$.\label{prop:predator-fixation}
\item $P_{(i,j)}(Z_n\to 0,\tZ_n\to \infty)>0$.\label{prop:non-iso-prey-fixation}
\end{enumerate}
\end{proposicion}

This result establishes that the fixation of the predator population is not possible because the mean growth rate of this species in absence of the prey population  is too small ($\rho_1\alpha m<1$). This model could therefore describe a predator-prey system where the prey is the main food source of predators, and consequently the survival of the predators is impossible if there is an unavailability of preys. The second part of the proposition establishes that the prey population has a positive probability of fixation. This is due to the fact that the chance of survival of the preys is obviously greater in the absence of predators and it only depends on their own reproductive capacity, which satisfies $\trho_2\talpha \tilde{m}>1$.

\begin{nota}\label{nota:fixation}
We emphasize that Proposition \ref{prop: prey and predator fixation}~\ref{prop:predator-fixation} is not in contradiction with situation \ref{sit:predators survival} because \ref{sit:predators survival} refers to the survival of each predator individual, while Proposition \ref{prop: prey and predator fixation} \ref{prop:predator-fixation} concerns the predator population. That is, it is possible that one predator survives in absence of preys, but in the long run, if there is no prey in the ecosystem the growth rate of predators is not enough to prevent the population from the extinction. \label{nota:fixation-sit predator survival}
\end{nota}


The following result on the ultimate extinction of the predator-prey system follows immediately by Proposition \ref{prop: prey and predator fixation}. 

\begin{corolario}\label{coro:extinction-conds}
Let $\{(Z_n,\tZ_n)\}_{n\in\N_0}$ be a PP-2SDDBPO. Then, $P_{(i,j)}(Z_n\to 0, \tZ_n\to 0)<1$ for any initial values $i,j>0$.
\end{corolario}

We recall that the extinction of both populations is always possible due to many reasons such as the possibility that no individual of both populations survives during the control phase, or that all the survivors of both species are of the same sex, which makes impossible to form new couples. In addition, there is a positive probability that the predator and prey couples produce no offspring if $p_0>0$ and $\tp_0>0$.

\section{Coexistence}\label{sec:coexistence}

We now examine the possibility of the coexistence of both species. Although until now, we have focussed on the fate of the total number of couples, $Z_n$ and $\tZ_n$, of each species, the results on coexistence in terms of the total number of individuals of each species before the control phase, $T_n$ and $\tT_n$, are equivalent in view of Lemma~\ref{lema:T=Z} in appendix. Thus, for our convenience in the proofs of the results in this section we establish the theorems in terms of $T_n$ and $\tT_n$.

Our results show that the coexistence of both species strongly depends on the parameter $\mu$. For that reason, we write
\begin{align}\label{Omega}
\Omega&=\Big\{\limsup_{n\to\infty} \frac{\tT_n}{T_n}\leq \mu\Big\}\cup \Big\{\liminf_{n\to\infty}\frac{\tT_n}{T_n}> \mu\Big\}\cup \Big\{\liminf_{n\to\infty}\frac{\tT_n}{T_n}\leq \mu< \limsup_{n\to\infty}\frac{\tT_n}{T_n}\Big\}.
\end{align}
%

\medskip

In the first theorem of this section we establish that the survival of none species is possible in the first event of \eqref{Omega}. As a consequence, the coexistence is also impossible for any initial number of individuals and regardless the values of the parameters. The idea is that if the number of preys per predator is not greater than $\mu$ from some generation on, then, on the one hand, the predator population will die out owing to the lack of resources to survive, and, on the other hand, more preys will be captured given the needs of the large number of predators compared to the prey population. 

\begin{teorema}\label{thm:coexistence-0}
Let $\{(Z_n,\tZ_n)\}_{n\in\N_0}$ be a PP-2SDDBPO. For any initial values $i,j>0$:
\begin{align*}
P_{(i,j)}\left(\Big\{\limsup_{n\to\infty} \frac{\tT_n}{T_n}\leq \mu\Big\} \cap \{T_n\to\infty\}\right)&=0,\\
P_{(i,j)}\left(\Big\{\limsup_{n\to\infty} \frac{\tT_n}{T_n}\leq \mu\Big\} \cap \{\tT_n\to\infty\}\right)&=0.
\end{align*}
In particular,
$$P_{(i,j)}\left(\Big\{\limsup_{n\to\infty} \frac{\tT_n}{T_n}\leq \mu\Big\} \cap \{T_n\to\infty,\tT_n\to\infty\}\right)=0.$$
\end{teorema}

\medskip

In our second result, we provide sufficient conditions for the coexistence of both species. This states that when $\trho_2\talpha \tm$ (the maximum  mean growth rate of female preys) is greater than $\rho_2\alpha m$ (the maximum mean growth rate of female predators), then the coexistence of both species is possible in the second event in \eqref{Omega}. This result was illustrated in Example 1 in Subsection \ref{subsec:illustrative-examples}.


\begin{teorema}\label{thm:coexistence-positive}
Let $\{(Z_n,\tZ_n)\}_{n\in\N_0}$ be a PP-2SDDBPO. If any of the following two conditions hold:
\begin{enumerate}[label=\emph{(\roman*)},ref=\emph{(\roman*)}]
\item $\rho_2\alpha m<\trho_2\talpha \tm$,
\item $\rho_2\alpha m=\trho_2\talpha \tm$, and  $\liminf_{x\to\infty} (\ts(x)\tm\talpha)/(s(x) m\alpha)>1$, 
\end{enumerate} 
then for any initial values $i,j>0$:
$$P_{(i,j)}\bigg(\Big\{\liminf_{n\to\infty}\frac{\tT_n}{T_n}> \mu\Big\}\cap \{T_n\to\infty,\tT_n\to\infty\}\bigg)>0.$$
In particular,
$$P_{(i,j)}(\{T_n\to\infty,\tT_n\to\infty\})>0.$$
\end{teorema}


%

\bigskip

%
%

The previous results do not cover the possibility of coexistence in the case $\rho_2\alpha m>\trho_2\talpha \tm$. Our numerical studies indicate that if the prey fixation does no occur at early generations of the process, then the process enters in an oscillation phase until the number of preys is zero, and then the ultimate extinction of the entire system occurs. In particular, in that phase the ratio $\tT_n/T_n$ fluctuates around $\mu$ (see Figure \ref{fig:path-density-sit2} in Example 2). We therefore expect that the coexistence is impossible on both events $\{\liminf_{n\to\infty} \tT_n/T_n>\mu\}$ and $\{\liminf_{n\to\infty}\tT_n/T_n\leq \mu< \limsup_{n\to\infty}\tT_n/T_n\}$ if $\rho_2\alpha m>\trho_2\talpha \tm$. Moreover, our simulation analysis suggests that the oscillation stage cannot occur forever, regardless the value of the parameters. We formalize those ideas in the following conjecture.


\begin{conjetura}\label{conj:coexistence-null}
Let $\{(Z_n,\tZ_n)\}_{n\in\N_0}$ be a PP-2SDDBPO. For any initial values $i,j>0$:
\begin{align*}
P_{(i,j)}\bigg(\Big\{\liminf_{n\to\infty}\frac{\tT_n}{T_n}\leq \mu< \limsup_{n\to\infty}\frac{\tT_n}{T_n}\Big\}\cap \{T_n\to\infty,\tT_n\to\infty\}\bigg)&=0.
\end{align*}
Moreover, if $\rho_2\alpha m>\trho_2\talpha \tm$, then for any initial values $i,j>0$:
\begin{align*}
P_{(i,j)}\bigg(\Big\{\liminf_{n\to\infty}\frac{\tT_n}{T_n}> \mu\Big\}\cap \{T_n\to\infty,\tT_n\to\infty\}\bigg)&=0,
\end{align*}
and in particular, 
$$P_{(i,j)}\big(T_n\to\infty,\tT_n\to\infty\big)=0.$$
\end{conjetura}

\begin{figure}[H]
\begin{center}
\includegraphics[width=0.32\textwidth]{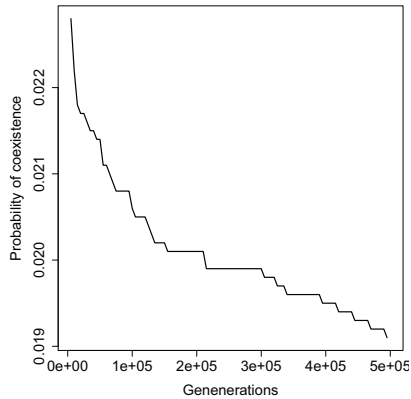}
\caption{Evolution of the probability $P(T_n>0,\tT_n>0)$ based on the simulation of $10^4$ processes following a simulated PP-2SDDBPO with $\rho_2\alpha m=1.176>\trho_2\talpha \tm=1.134$, and $\mu=2$.}\label{fig:probability_coexistence}
\end{center}
\end{figure}

To support this conjecture we performed a simulation of $10^4$ processes until generation $5\cdot 10^5$, with the same parameters as in Example 2 in Subsection~\ref{subsec:illustrative-examples}. We estimated the probability of coexistence of both species by using the proportion of processes, among the $10^4~$ simulated ones, where both species have survived. We show its evolution over the generations in Figure~\ref{fig:probability_coexistence},  where we observe how that probability decreases to zero as the number of generations increases.

\section{Modelling of the population of wolves and moose in Isle Royale}\label{sec:real-data}

In this section, we evaluate the suitability of the proposed model to describe the evolution of the number of wolves and moose studied in Isle Royale National Park over 61 years (see \cite{Vucetich2012}). The Isle Royal National Park is an international biosphere reserve located in Lake Superior (Michigan, USA). The wolf and moose populations in this ecosystem are well known since their study started back in 1959. Indeed, it is one of the predator-prey system with the longest continuous study in the world. In the following, we show that this real dataset can be easily mimicked using the model presented in this paper. To that end, we generated a path of a PP-2SDDBPO where the values of the parameters are reasonable bearing in mind what one can observe in populations of wolves and moose. We remark that our aim here is to illustrate the adjustment and usefulness of our process to model the real world situations, but making inference on the parameters of the model exceeds the scope of this paper.

First, we note that the reproduction, control and mating phases described in the Section~\ref{sec:Definition} faithfully adjust to the evolution of the populations of wolves and moose. This is owing to the fact that both species have a sexual reproduction with promiscuous mating where not all females are able to reproduce (see, for example, \cite{Ballenberghe-1993} for the moose population or \cite{Peterson-2002} for the wolf population).

Next, we describe the choice of the parameter values and their relationship with the populations. We shall start with the reproduction laws. On the one hand, given that several authors agreed that there is an unbalanced sex ratio in wolves, with more males than females in the population (see \cite{Cowan-1947}, \cite{Mech-1975} or \cite{Stenlund-1955}), we considered $\alpha=0.35$ as sex ration for the predator. On the other hand, we took $\talpha=0.49$ as the sex ratio for the preys motivated by the fact some studies claimed that there is no statistically significant difference between the population of males and of females in moose populations (see \cite{Peek-Urich-Mackie-1976} or \cite{Shubin-Yazan-1959}).

Regarding the reproduction, since a typical wolf litter consists of four to seven pups, for the mean of the predator reproduction law we fixed $m=4$. Similarly, we took $\tm=2.5$ as mean of the prey reproduction law because moose females have around two embryos per pregnancy each year (see \cite{Danilkin-1998}).

Additionally, we selected $\rho_1=\trho_1=0.05$, both of them very close to 0. The first choice is due to the fact that the Isle Royale is an isolated place where the moose is the main food source for the wolves. Thus, their survival is very difficult in absence of moose (see \cite{Peterson-Page-1988}). The latter is because the probability that the moose could avoid predators is small if the proportion of wolves per prey goes to infinity. For the probabilities $\rho_2$ and $\trho_2$, we considered $\rho_2=0.8$ and $\trho_2=0.93$. The choice $\rho_2=0.8$ is motivated by the fact that predators could die by natural reasons despite having enough preys in the population. The value $\trho_2=0.93$ is justified by several facts. First, because wolves are their main (or even their unique) predator; second, because moose hunting is prohibited in the island, and third, because the possibility of immigration and emigration is negligible (see \cite{Vucetich-Peterson-2004}). We note that the choice of the parameters is reasonable in the context of Isle Royale biosphere, but they also satisfy conditions \eqref{eq: rho1alfam<1<rho2alfam} and $\rho_2\alpha m < \trho_2\talpha \tm$.

\begin{figure}[H]
\centering\includegraphics[width=0.32\textwidth]{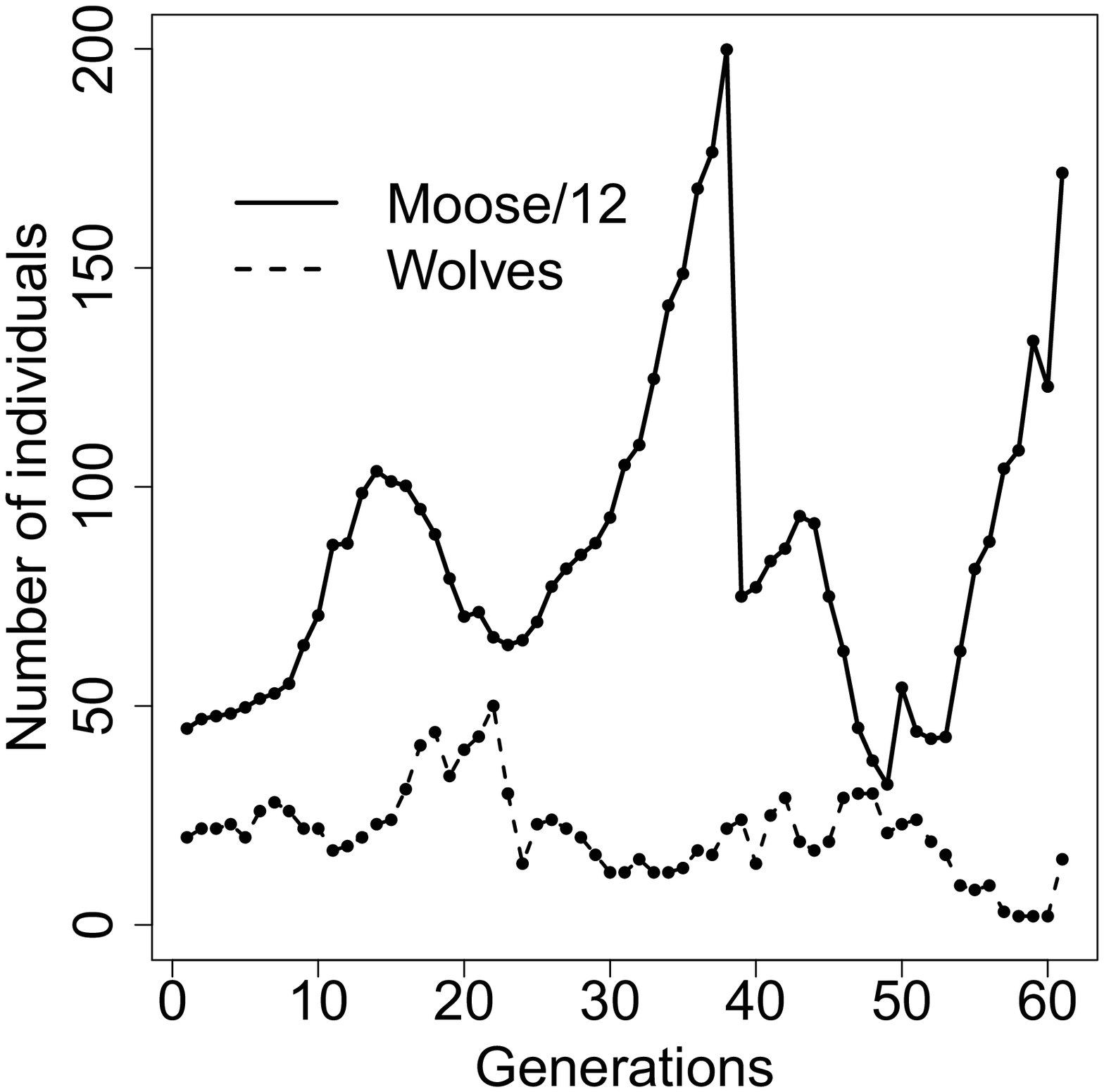}
\includegraphics[width=0.32\textwidth]{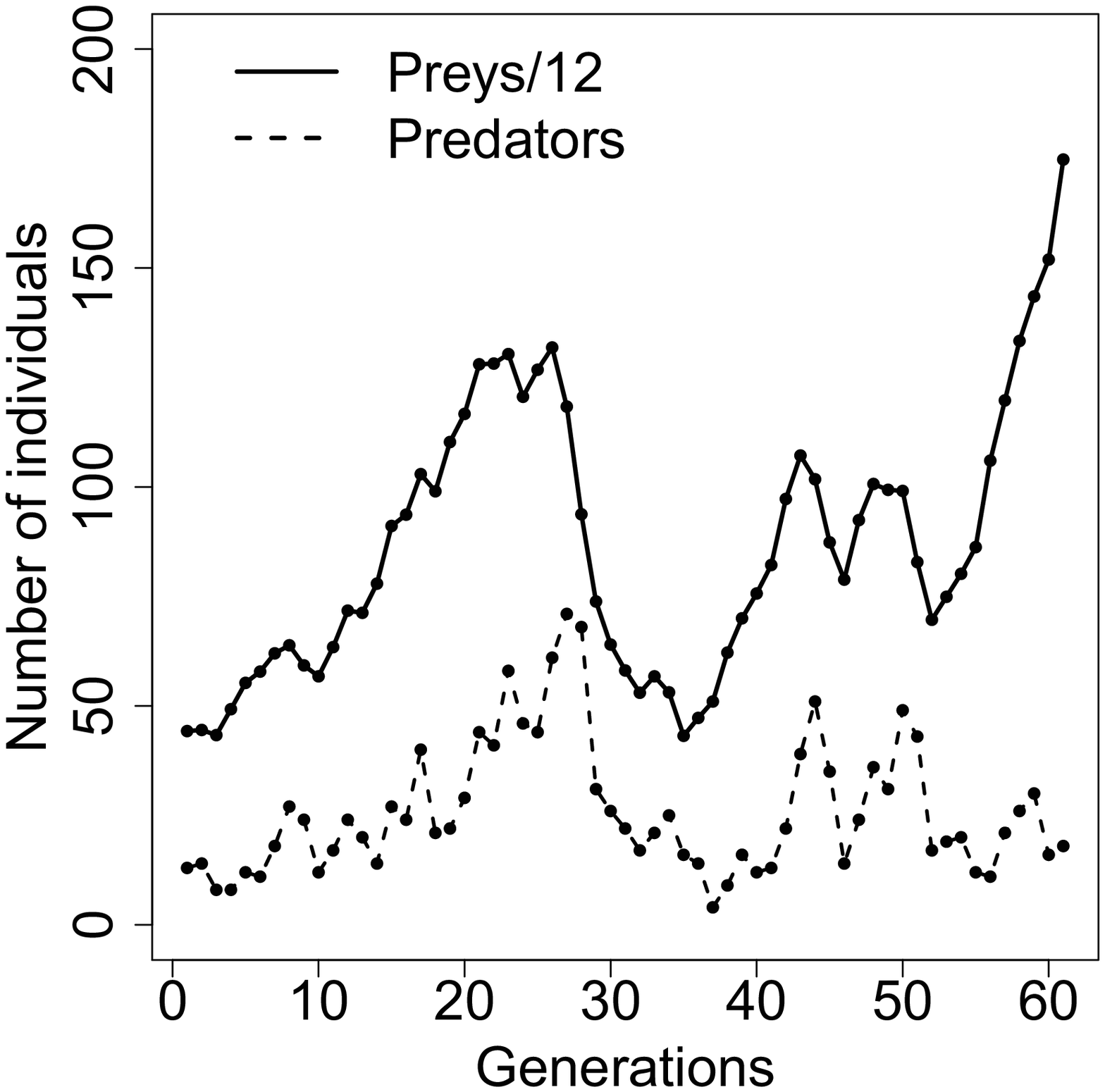}\includegraphics[width=0.32\textwidth]{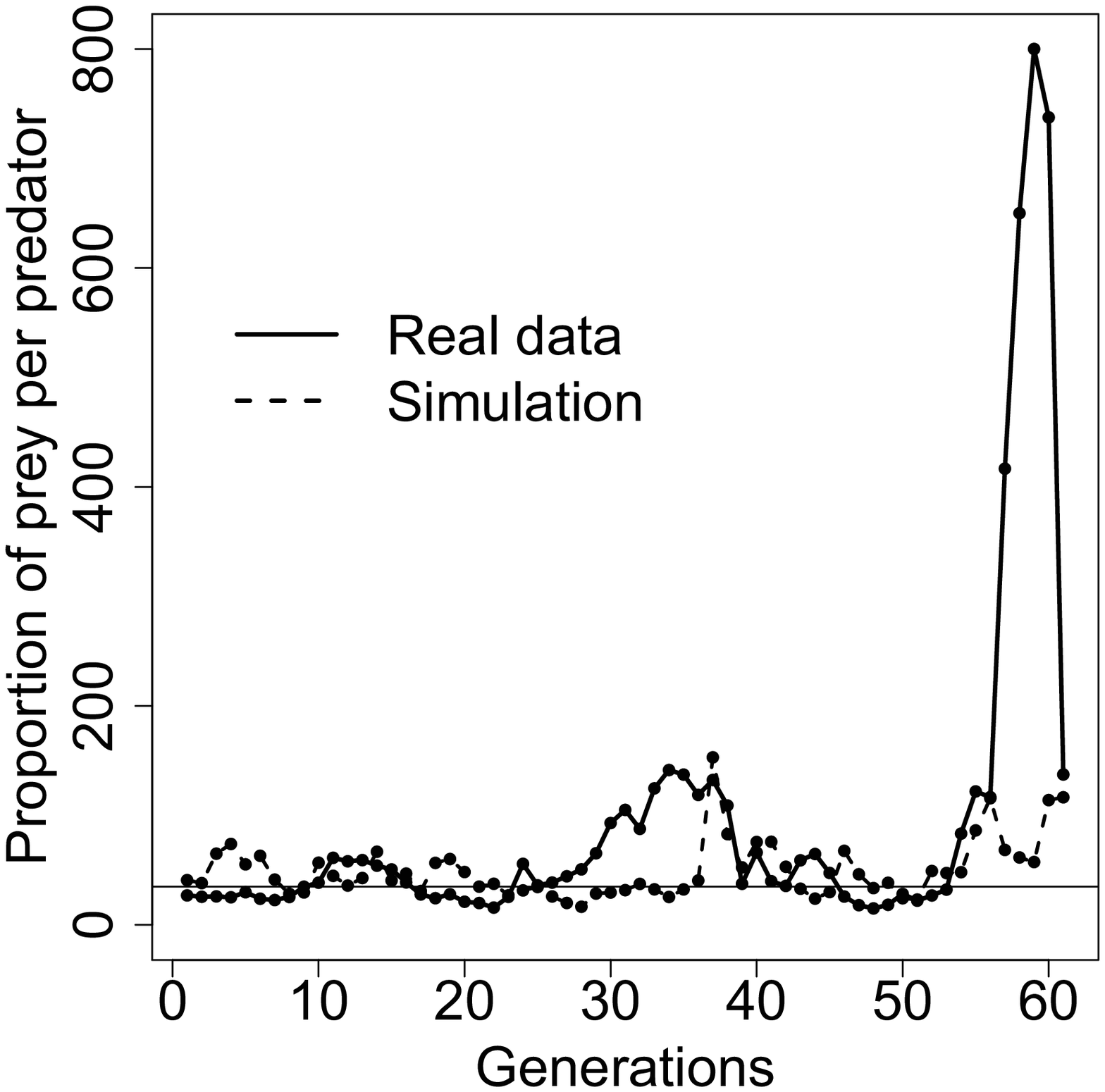}
\caption{Left: evolution of the number of wolves (dashed line) and moose (solid line) from 1959 to 2019 in the real dataset. Centre: evolution of the number of predators (dashed line) and of the number of preys (solid line) previous to the control phase over 61 generations in the simulated process. Right: proportion of preys per predator over the generations for both the real data (solid line) and the simulated path (dashed line). The horizontal line represents the value $\mu=35$.}\label{fig:real-path}
\end{figure}

Finally, given that our simulation studies show that the proportion of preys per predator oscillates around the parameter $\mu$ in the present model, we therefore considered that $\mu=35$ is a plausible value as illustrated in Figure \ref{fig:real-path} (right). For the PP-2SDDBPO with these parameters we simulated 61 generations starting with the same number of individuals as the real dataset, that is, 20 predators and 538 preys. We remark that neither the initial number of couples nor their offspring are known for this real  predator-prey system. Moreover, we fixed the function $f_1(\cdot)$ in Remark \ref{nota:descrip-def} to define the probabilities of survival ($s(\cdot)$ and $\ts(\cdot)$). 

First, in Figure \ref{fig:real-path} (left) we plot the total number of wolves and the total number of dozens of moose observed in the Isle Royale National Park from 1959 to 2019. Next, in Figure \ref{fig:real-path} (centre) we show the evolution of the total number of predators and the total number of dozens of preys of the simulated path in each generation. To obtain this last path, we simulated processes following the model described above, and we found that it is easy to get paths with a similar behaviour to the one showed in Figure \ref{fig:real-path}. Note that the total number of individuals which is observed in the population is previous to the control phase, that is, they correspond to the values of the variables $\{(T_n,\tT_n): \ n=1,\ldots,61\}$ in our model. Some of these individuals will participate in the mating phase but, initially, we do not have any knowledge about how many of them.  The total number of the individuals after the control, that is, $\{(\phi_n(T_n,\tT_n),\tphi_n(T_n,\tT_n)): \ n=1,\ldots,61\}$, is therefore unknown.

By comparing both graphs one can observe a similar behaviour. Overall, the total number of dozens of preys is always greater than the total number of predators over the generations for both the real and the simulated dataset. Moreover, the ratio between the number of individuals in both populations remains oscillating around some constant value over the generations. This fact is an indicator that the model can capture correctly the   evolution of the proportion of predators per prey, as can be seen in Figure~\ref{fig:real-path} (right). In relation to the cyclic behaviour of our model, one can notice that stages of growth in the number of preys are followed by stages of growth in the predator population and vice-versa, as it occurs with the real data. Moreover, when the number of preys drops, the same tendency is appreciated in the number of predators in the next generations. 

It is worth to mention that the real dataset presents a dramatic drop in moose population in the mid 1990s, from a total of 2398 in 1996 to 900 in 1997. Some authors claimed that this sharp drop in the real data might have been caused by the severe climatic conditions during that period (see \cite{Vucetich-Peterson-2004} or \cite{Montgomery-2013}). We remark that our model is not able to capture this type of extreme changes because the offspring distributions do not vary over the generations, and then, the drops and rises of our model are more moderated. Nevertheless, one could think of including a varying environment for the reproduction laws to describe these situations, however, we do not analyse this fact in depth because it is beyond the scope of this paper.


\section{Discussion}\label{sec:Discussion}

The aim of this work is to model the generation-by-generation evolution of populations of predators and preys as well as their interaction when both of them have sexual reproduction. For that purpose, a two-type two-sex branching process with a control in the number of individuals of each species is introduced in the present paper.

As usual in two-sex branching processes theory, the reproduction and mating phases are taking into account in the definition of the model. Additionally, a control stage is also considered to model the relationship between the species. Those three stages are repeated in each generation. The description for a given generation is the following: first, in the reproduction phase, a random number of offspring is produced by couples of each species. Next, due to the interaction between predators and preys, some of these individuals could die. This is modelled in the control phase by means of binomial distributions where the probability of success for each species depends on the density of preys per predator. To determine how many survivor individuals are females and males multinomial distributions are considered. In this final mating phase, couples are formed via a promiscuous mating in both species. These couples produce offspring in the next generation and this three-stage cycle is repeated again.

We have proved some results concerning with the fate of the species in the population (extinction, fixation and coexistence). In particular, we have established that for any initial number of couples of both species the probability of fixation of the predator population is null due to the fact that the mean growth rate of this species is less than one in absence of preys. Additionally, the prey population has a positive probability of fixation because its mean growth rate is greater than one in absence of predators, and as a consequence, the probability of extinction is less than one. 

The study of the possibility of the coexistence is more complex due to its dependence on the parameter $\mu$. This parameter represents the necessary prey density per predator so that both populations remain stable. Specifically, we have established that the coexistence is not possible if the supremum of the density of preys per predator is not above $\mu$ in the long run. Furthermore, when the infimum of such a density is eventually greater than $\mu$, we provide a sufficient condition for the coexistence to have a positive probability. Based on a simulated study, we have also conjectured that, on the one hand, the oscillation phase  cannot last forever on the coexistence event. On the other hand, when the maximum mean growth rate of female predators is greater than the maximum mean growth rate of female preys, and both populations are still alive, the process fluctuates during long time until its final extinction and then the coexistence of both species is not possible.

The main novelty of this model regarding to the one introduced in \cite{GutierrezMinuesa2020} for population with sexual reproduction and also with respect to other predator-prey models in the context of branching process (see \cite{Alsmeyer-1993} or \cite{Coffey-Buhler-1991}) is that under certain conditions this process presents the typical fluctuations or oscillations which appear in nature in the majority of predator-prey systems (see, for example, \cite{Gilpin1973}). In those oscillations we observe periods with a large number of predators and preys which are followed by periods of a smaller number of predators and preys. Two elements are the key of this model to obtain this behaviour: first, the introduction of the density of preys per predator in the survival function in the control phase of each species and second, the introduction of the parameter $\mu$ representing the necessary proportion of preys per predator to maintain both populations balanced.

As a final conclusion, we emphasize that this model has shown to be appropriate to fit real predator-prey ecosystems. The fact of considering sexual reproduction between the individuals of species results particularly useful to model the interaction of mammals populations of predators and preys.


 

\section*{Funding}


This research was funded by the Ministerio de Ciencia e Innovaci\'on [grant PID2019-108211GBI00 /AEI/10.13039/501100011033].

\appendix
\section*{Appendix}

\begin{lema}\label{lema:T=Z}
Let $\{(Z_n,\tZ_n)\}_{n\in\N_0}$ be a PP-2SDDBP. Then:
$$\{T_n\to\infty,\tT_n\to\infty\}=\{Z_n\to\infty,\tZ_n\to\infty\}\quad a.s.$$
\end{lema}


\begin{Prf}[Theorem \ref{thm:coexistence-0}]
We shall prove the first part. The second one is obtained with the same arguments. We follow the same steps as in the proof of Proposition~10 in \cite{GutierrezMinuesa2020}; we provide the details for the readers convenience. First, note that 
$$\left\{\limsup_{n\to\infty} \frac{\tT_n}{T_n}\leq \mu\right\}\subseteq \bigcup_{N=1}^\infty \left\{\sup_{n\geq N} \frac{\tT_n}{T_n}\leq \mu\right\},$$
and then 
it is enough to prove that for every $N>0$,
\begin{equation}\label{equ:prob-sup}
P_{(i,j)}\left(\left\{\sup_{n\geq N} \frac{\tT_n}{T_n}\leq \mu\right\}\bigcap \Big\{T_n\to\infty\Big\}\right)=0.
\end{equation}

Next, observe that for all $n\in\N_0$, using Proposition \ref{prop:expectations}~\ref{prop:expectations-ii} we have
\begin{equation}\label{equ:mart-Z}
E[T_{n+1}|\mathcal{G}_n]\leq \alpha m T_n s(\tT_n/T_n)\leq T_n \ \mbox{ a.s.\quad on } \{\tT_n/T_n\leq \mu\}.
\end{equation}

Now, let us fix $N>0$, and introduce the sequence of r.v.s $\{X_n\}_{n\in\N_0}$ defined as:
$$X_n=\begin{cases}
T_{N+n}, & \mbox{ if } N+n\leq \tau(\mu),\\
T_{\tau(\mu)}, & \mbox{ if } N+n>\tau(\mu),\end{cases}\qquad n\in\N_0,$$
where $\tau(\mu)$ is the stopping time:
$$\tau(\mu)=
\begin{cases}
\infty, & \mbox{ if } \sup_{n\geq N}\frac{\tT_n}{T_n} \leq \mu,\\
\min\big\{n\geq N: \frac{\tT_{n}}{T_{n}} > \mu\big\}, & \mbox{ otherwise}.
\end{cases}$$

If we prove that $\{X_n\}_{n\in\N_0}$ is a non-negative super-martingale with respect to $\{\G_{N+n}\}_{n\in\N_0}$, then applying the martingale convergence theorem, we have that the sequence $\{X_n\}_{n\in\N_0}$ converges almost surely to a non-negative and finite limit $X_\infty$, where
$$X_{\infty}=\begin{cases}
\lim_{n\to\infty}T_n, & \mbox{ if } \sup_{n\geq N} \frac{\tT_n}{T_n}\leq \mu,\\ 
T_{\tau(\mu)}, & \mbox{ otherwise.}
\end{cases}$$
As a consequence, \eqref{equ:prob-sup} is obtained and the proof of the first part finishes. 


To prove that $\{X_n\}_{n\in\N_0}$ is a super-martingale with respect to $\{\G_{N+n}\}_{n\in\N_0}$, we first note that the variable $X_n$ is $\G_{N+n}$-measurable for any $n\in\N_0$. Next, we show that $E[X_{n+1}|\G_{N+n}]\leq X_n$ a.s. for each $n\in\N_0$.

Let us fix $n\in\N_0$. Then, we have that if $\frac{\tT_{N+k}}{T_{N+k}}\leq \mu$, for every $k=0,\ldots,n$, then the stopping time satisfies $\tau(\mu)\geq N+n+1$, and using \eqref{equ:mart-Z} we get
$$E[X_{n+1}|\G_{N+n}]=E\left[T_{N+n+1}|\G_{N+n}\right]\leq T_{N+n}=X_n \ \mbox{a.s.\quad on } \bigcap_{k=0}^n\left\{\frac{\tT_{N+k}}{T_{N+k}}\leq \mu\right\}.$$
Now, if there exists $k\in\{1,\ldots,n\}$ satisfying $\frac{\tT_{N}}{T_{N}}\leq \mu,\ldots,\frac{\tT_{N+k-1}}{T_{N+k-1}}\leq \mu$ and $\frac{\tT_{N+k}}{T_{N+k}}>\mu$, then for the stopping time we have $\tau(\mu)= N+k\leq N+n< N+n+1$, and consequently, for the conditional expectation we get
$$E[X_{n+1}|\G_{N+n}]=E[T_{\tau(\mu)}|\G_{N+n}]=T_{\tau(\mu)}=X_n \ \mbox{a.s.\quad on } \bigcap_{l=0}^{k-1}\left\{\frac{\tT_{N+l}}{T_{N+l}}\leq \mu\right\}\bigcap\left\{\frac{\tT_{N+k}}{T_{N+k}}>\mu\right\}.$$
Finally, for all $n\in\N_0$, the stopping time is $\tau(\mu)=N<N+n+1$ if $\frac{\tT_{N}}{T_{N}}>\mu$, and, consequently,
$$E[X_{n+1}|\G_{N+n}]=E[T_{N}|\G_{N+n}]=T_{\tau(\mu)}=X_n \ \mbox{a.s.\quad on } \left\{\frac{\tT_{N}}{T_{N}}>\mu\right\}.$$

Combining all the above, if $A_n=\cap_{k=0}^n\big\{\frac{\tT_{N+k}}{T_{N+k}}\leq \mu\}$ is a measurable set with respect to the $\sigma$-algebra $\G_{N+n}$, then we get
\begin{align*}
E[X_{n+1}|\G_{N+n}]
&= E[X_{n+1}|\G_{N+n}]\indi{A_n}+E[X_{n+1}|\G_{N+n}]\indi{A_n^c}\leq T_{N+n}\indi{A_n} + T_{\tau(\mu)}\indi{A_n^c}=X_n \mbox{ a.s.}
\end{align*}

\end{Prf}

\vspace{2ex}

In order to prove Theorem~\ref{thm:coexistence-positive}, we make use of the following easy-to-prove lemma.



%

\vspace{2ex}

\begin{lema}\label{prop:states}
Let $\{(Z_n,\tZ_n)\}_{n\in\N_0}$ be a PP-2SDDBP. If $p_0+p_1+p_2<1$ and $\tilde{p}_0+\tilde{p}_1+\tilde{p}_2<1$, then the sets $\{(i,0): i>0\}$, $\{(0,j): j>0\}$ and $\{(i,j): i,j>0\}$ are classes of communicating states and each state leads to the state $(0,0)$. Furthermore, the process can move from the last set to the others in one step.
\end{lema}

\vspace{2ex}

\begin{Prf}[Theorem \ref{thm:coexistence-positive}]
The proof follows similar arguments to those in Theorem 3 of \cite{GutierrezMinuesa2020}. Let us fix $N\in\N$. We shall prove that we can take $I_0,J_0\in\N$ sufficiently large so that 
\begin{equation}\label{eq:thesis-thm:coexistence-positive}
P_{(i,j)}\bigg(\Big\{\inf_{n\geq N} \frac{\tT_n}{T_n}>\mu\Big\}\cap \{T_n\to\infty,\tT_n\to\infty\}\bigg)>0,\quad \text{ for each }i,j\in\N,\ i\geq I_0,\ j\geq J_0.
\end{equation}
If this holds, then the result also holds for any $i,j\in\N$. Indeed, if $i,j\in\N$ satisfy that either $i<I_0$ or $j<J_0$, then the existence of $n_0\in\N$ such that 
$$P(Z_{n_0}=I_{0},\tZ_{n_0}=J_{0}|Z_0=i,\tZ_0=j)>0,$$
is guaranteed by Lemma~\ref{prop:states}, and using the Markov property we have 
\begin{align*}
P_{(i,j)}&\bigg(\Big\{\liminf_{n\to\infty} \frac{\tT_n}{T_n}>\mu\Big\}\cap \{T_n\to\infty,\tT_n\to\infty\}\bigg)\geq\\
&\geq P_{(i,j)}\bigg(\Big\{\inf_{n\geq N+n_0} \frac{\tT_n}{T_n}>\mu\Big\}\cap \{T_n\to\infty,\tT_n\to\infty\}\bigg)\\
&\geq 
P_{(I_0,J_0)}\Big(\inf_{n\geq N} \frac{\tT_n}{T_n}>\mu, T_n\to\infty,\tT_n\to\infty\Big)\cdot P_{(i,j)}(Z_{n_0}=I_0,\tZ_{n_0}=J_0)>0.
\end{align*}


Hence, it only remains to prove \eqref{eq:thesis-thm:coexistence-positive}.  
We shall start with the case $\trho_2\talpha \tm>\rho_2\alpha m>1$. Then, we can choose $\varepsilon_1,\varepsilon_2>0$ such that 
$$1<\rho_2\alpha m-\varepsilon_1=\eta_1<\trho_2\talpha \tm-\varepsilon_2=\eta_2.$$
Bearing in mind that $p_0+p_1+p_2<1$ and $\tilde{p}_0+\tilde{p}_1+\tilde{p}_2<1$, we can also take $r_1,r_2\geq 3$ such that $p_{r_1}>0$ and $\tilde{p}_{r_2}>0$, and then, $\eta_0=r_1-1>1$ and $\tilde{\eta}_0=r_2-1>1$. Moreover, since $\lim_{x\to \infty} s(x)=\rho_2$, and $\lim_{x\to \infty} \ts(x)=\trho_2$, there exists $M>\mu$ such that $|\alpha m\rho_2-\alpha m s(x)|\leq\varepsilon_1/4$, and $|\talpha \tm\trho_2-\talpha \tm \ts(x)|\leq\varepsilon_2/4$, for every $x\geq M$. 


We now consider the next events:
$$A_0=\{\eta_0Z_0<T_1, \tilde{\eta}_0\tZ_0<\tT_1\}\quad \mbox{ and }\quad A_n=\{\eta_1T_n<T_{n+1},\eta_2\tT_n<\tT_{n+1},M T_n\leq \tT_n\},\quad n\in\N.$$


Then, we have that
\begin{align*}\label{equ: non-aisolated coexistence probAn1}
P_{(i,j)}\Big(\inf_{n\geq N}\frac{\tT_n}{T_n}> \mu, T_n\to\infty,\tT_n\to\infty\Big)&\geq P_{(i,j)}\Big(\inf_{n\in\N}\frac{\tT_n}{T_n}> \mu, T_n\to\infty,\tT_n\to\infty\Big)\\
&\geq P_{(i,j)}\left(\cap_{n=0}^{\infty}A_n\right)\\
&=\lim_{n\to\infty}P_{(i,j)}\left(\cap_{l=0}^{n}A_l\right)\\
&=\lim_{n\to\infty}P_{(i,j)}(A_0)\prod_{l=1}^{n}P\left(A_l|\cap_{k=0}^{l-1}A_k\cap \{Z_0=i,\tZ_0=j\}\right).
\end{align*}

Let us also introduce the sets 
$$B_l=\{(x,y)\in\N^2: i\eta_0\eta_1^{l-1}<x,\ j\tilde{\eta}_0\eta_2^{l-1}<y,\ Mx\leq y\},\quad \text{ for each }l\in\N,$$
and observe that
the family 
$$\cap_{k=0}^{l-1}A_k\cap \{(T_l,\tT_l)=(x,y)\}\cap \{Z_0=i,\tZ_0=j\},\quad x,y\in\N,$$
is a partition of the set $\cap_{k=0}^{l-1}A_k \cap \{Z_0=i,\tZ_0=j\}$. Then, using the Markov property it is not difficult to deduce that
\begin{align*}
P\big(A_l|\cap_{k=0}^{l-1}A_k&\cap \{Z_0=i,\tZ_0=j\}\big)\\
&\geq \inf_{(x,y)\in B_l} P\left(A_{l}|\cap_{k=0}^{l-1}A_k\cap \{(T_l,\tT_l)=(x,y)\}\cap \{Z_0=i,\tZ_0=j\}\right)\nonumber\\
&=\inf_{(x,y)\in B_l} P\left(A_{l}|(T_l,\tT_l)=(x,y)\right)\nonumber\\
&=\inf_{(x,y)\in B_l} P(A_1|(T_1,\tT_1)=(x,y)).
\end{align*}


To finish the proof, we start bounding the probability $P(A_1^c|(T_1,\tT_1)=(x,y))$ from above when $(x,y)\in B_l$. 
Observe that if $(x,y)\in B_l$, then $P\big(\tT_1/T_1<M|(T_1,\tT_1)=(x,y)\big)=0$. Next, using Proposition \ref{prop:expectations}~\ref{prop:expectations-i} we have that for $x\in\N$, 
\begin{align*}
P\big(T_2\leq \eta_1 T_1 & |(T_1,\tT_1)=(x,y)\big)=\\
&=P\Big(x\big(\varepsilon_1-\alpha m \rho_2+\alpha ms(y/x)-\alpha m s(y/x)(1-s(y/x)(1-\alpha))^{x-1}\big)\\
&\phantom{=}\hspace{2em}\leq E[T_2|(T_1,\tT_1)=(x,y)]-T_2 \big|(T_1,\tT_1)=(x,y)\Big).
\end{align*}

On the one hand, by the properties of the function $s(\cdot)$, we have  
$$\alpha m s(y/x)(1-s(y/x)(1-\alpha))^{x-1}\leq \alpha m\rho_2(1-\rho_1(1-\alpha))^{x-1},$$
and since $0<1-\rho_1(1-\alpha)<1$, there exists $I_0\in\N$ satisfying that
$$\alpha m s(y/x)(1-s(y/x)(1-\alpha))^{x-1}<\varepsilon_1/4,\quad \text{ for every }x\geq I_0, y\in\N.$$

On the other hand, $|\alpha m\rho_2-\alpha m s(y/x)|\leq \varepsilon_1/4$, for every $x,y\in\N$, with $y\geq M x$.
Thus, combining all the above, for each $x\geq I_0$ and $y\geq M x$, by Chebyschev's inequality we have
\begin{align*}
P\big(T_2\leq \eta_1 T_1&|(T_1,\tT_1)=(x,y)\big)\leq\\
&\leq P\Big(x\big(\varepsilon_1-\alpha m \rho_2+\alpha ms(y/x)-\alpha m s(y/x)(1-s(y/x)(1-\alpha))^{x-1}\big)\\
&\phantom{=}\hspace{2em}\leq |E[T_2|(T_1,\tT_1)=(x,y)]-T_2| \big|(T_1,\tT_1)=(x,y)\Big)\\
&\leq P\Big(\frac{x \varepsilon_1}{2}\leq |E[T_2|(T_1,\tT_1)=(x,y)]-T_2| \big|(T_1,\tT_1)=(x,y)\Big)\\
&\leq \frac{4 Var[T_2|(T_1,\tT_1)=(x,y)]}{\varepsilon_1^2 x^2}.
\end{align*}
Taking into account Proposition \ref{prop:expectations}~\ref{prop:expectations-ii} we can choose two constants $C_1>0$ and $C_2>0$ such that for $x\geq I_0$ and $y\geq M x$,
\begin{align*}
P\big(T_2\leq \eta_1 T_1|(T_1,\tT_1)=(x,y)\big)
&\leq \frac{C_1}{x}+C_2(1-\rho_1(1-\alpha))^{x}.
\end{align*}

Using the same arguments, we can take $J_0\geq M I_0$ such that for $y\geq J_0$, $x\in\N$, with $y/x\geq M$, we have that 
\begin{align*}
P\big(\tT_2\leq \eta_2 \tT_1&|(T_1,\tT_1)=(x,y)\big)\leq\\
&\leq P\Big(y\big(\varepsilon_2-\talpha \tm \trho_2+\talpha \tm\ts(y/x)-\talpha \tm \ts(y/x)(1-\ts(y/x)+\talpha \ts(y/x))^{y-1}\big)\\
&\phantom{=}\hspace{2em}\leq  |E[\tT_2|(T_1,\tT_1)=(x,y)]-\tT_2| \big|(T_1,\tT_1)=(x,y)\Big)\\
&\leq P\Big(\frac{y \varepsilon_2}{2}\leq |E[\tT_2|(T_1,\tT_1)=(x,y)]-T_2| \big|(T_1,\tT_1)=(x,y)\Big)\\
&\leq \frac{C_3}{y}+C_4(1-(1-\talpha)\trho_1)^{y},
\end{align*}
for some constants $C_3>0$ and $C_4>0$.

Now, to bound the probability $P_{(i,j)}(A_0)$ from below observe that
\begin{align*}
P_{(i,j)}(A_0)=P_{(i,j)}(T_1>\eta_0 Z_0,\tT_1>\tilde{\eta}_0 \tZ_0)
&\geq P(T_1=r_1 i|Z_0=i)\cdot P(\tT_1=r_2 j|\tZ_0=j)
=p_{r_1}^i\tilde{p}_{r_2}^j>0,
\end{align*}
where we have used that $\lfloor\eta_0 i\rfloor+1\leq r_1 i$ and $\lfloor\tilde{\eta}_0 i\rfloor+1\leq r_2 i$, for any $i\in\N$, and $\lfloor \cdot \rfloor$ denotes the integer part function.

Finally, if $i\geq I_0$ and $j\geq J_0$, then since $\eta_0,\tilde{\eta}_0,\eta_1,\eta_2>1$, we have $i\eta_0\eta_1^{l-1}\geq I_0$ and $j\tilde{\eta}_0\eta_2^{l-1}\geq J_0$, for all $l\in\N_0$. Now, it is not difficult to verify that
\begin{align*}
P_{(i,j)}&(T_n\to\infty,\tT_n\to\infty) \geq 
p_{r_1}^i \tilde{p}_{r_2}^j \prod_{l=1}^{\infty}\inf_{(x,y)\in B_l}P(A_1|(T_1,\tT_1)=(x,y))\\
&\geq p_{r_1}^i \tilde{p}_{r_2}^j \cdot\prod_{l=1}^\infty \inf_{(x,y)\in B_l}\bigg(1-\frac{C_1}{x}-C_2(1-(1-\alpha)\rho_1)^{x}-\frac{C_3}{y}-C_4(1-(1-\talpha)\trho_1)^{y}\bigg)\\
&\geq p_{r_1}^i \tilde{p}_{r_2}^j \cdot\prod_{l=1}^\infty\bigg(1-\frac{C_1}{i\eta_0\eta_1^{l-1}}-C_2(1-(1-\alpha)\rho_1)^{i\eta_0\eta_1^{l-1}}-\frac{C_3}{j\tilde{\eta}_0\eta_2^{l-1}}\\
&\phantom{\geq} -C_4(1-(1-\talpha)\trho_1)^{j\tilde{\eta}_0\eta_2^{l-1}}\bigg)>0.
\end{align*}

\vspace{2ex}

Now we turn to the case $\trho_2\tm\talpha=\rho_2 m\alpha$, and $\liminf_{x\to\infty} (\ts(x)\tm\talpha)/(s(x) m\alpha)>1$. Then, we can choose $M_0>\mu$ such that $\ts(x)\tm\talpha>s(x) m\alpha$ for every $x\geq M_0$. Let $\eta_1=s(M_0) m\alpha>1$ and $\varepsilon_1=\rho_2 m\alpha-\eta_1>0$. Moreover, there exists $M'>\mu$ such that $|\rho_2 m\alpha-s(x) m\alpha|\leq \varepsilon_1/4$, for every $x\geq M'$.

Next, we take $\eta_2=\ts(M_0) \tm\talpha$, and $\varepsilon_2=\trho_2 \tm\talpha-\eta_2>0$, which satisfies $\eta_2>\eta_1>1$. Moreover, there exists $M>M'$ such that $|\trho_2 \tm\talpha-\ts(x) \tm\talpha|\leq \varepsilon_2/4$, for every $x\geq M$. The proof continues with the same arguments as before and the result follows.
\end{Prf}

\vspace{2ex}
%
%
%
%
%


\end{document}